\documentclass[fleqn,usenatbib]{mnras}

\usepackage{newtxtext,newtxmath}

\usepackage[T1]{fontenc}
\usepackage{ae,aecompl}
\usepackage{xcolor}

\usepackage{graphicx}	
\usepackage{amsmath}	
\usepackage{amssymb}	
\usepackage{mathtools}
\usepackage{bm}
\usepackage{multirow}

\newcommand{\mrm}[1]{\mathrm{#1}} 

\title[CMB beam convolution]{Full-sky beam convolution for cosmic microwave background applications}
\author[Adriaan~J.~Duivenvoorden et al.]{
Adriaan~J.~Duivenvoorden,$^{1}$\thanks{E-mail: adri.duivenvoorden@fysik.su.se}
Jon~E.~Gudmundsson,$^{1}$ Alexandra~S.~Rahlin$^{2,3}$
\\
$^{1}$The Oskar Klein Centre for Cosmoparticle Physics,
Department of Physics, Stockholm University, SE-106 91 Stockholm, Sweden\\
$^{2}$Fermi National Accelerator Laboratory, Batavia, IL, USA\\
$^{3}$Kavli Institute for Cosmological Physics, University of Chicago, Chicago, IL, USA
}

\date{Accepted XXX. Received YYY; in original form ZZZ}

\pubyear{2018}

\begin{document}
\label{firstpage}
\pagerange{\pageref{firstpage}--\pageref{lastpage}}
\maketitle

\begin{abstract}
We introduce a publicly available full-sky beam convolution code library intended to inform the design of future cosmic microwave background (CMB) instruments and help current experiments probe potential systematic effects. 
The code can be used to assess the impact of optical systematics on all stages of data reduction for a realistic experiment, including analyses beyond power spectrum estimation, by generating signal timelines that may serve as input to full analysis pipelines.
The design and mathematical framework of the \texttt{Python} code is discussed along with a few simple benchmarking results. 
We present a simple two-lens refracting telescope design and use it together with the code to simulate a year-long dataset for 400 detectors scanning the sky on a satellite instrument.
 The simulation results identify a number of sub-leading optical non-idealities and demonstrate significant $B$-mode residuals caused by extended sidelobes that are sensitive to polarized radiation from the Galaxy. 
 For the proposed design and satellite scanning strategy, we show that a full physical optics beam model generates $B$-mode systematics that differ significantly from the simpler elliptical Gaussian model. The code is available at \href{https://github.com/adrijd/beamconv}{https://github.com/adrijd/beamconv}.

\end{abstract}




\section{Introduction}
\label{sec:introduction}

Many current cosmological observing programs are focused on a conjectured imprint of primordial gravitational waves in the degree-scale polarization anisotropies of the cosmic microwave background (CMB). This, together with efforts to quantify CMB polarization on both small (arcmin) as well as the largest possible angular scales, are driving a significant increase in the sensitivity of CMB instruments \citep{Abazajian2015, Litebird2016, Abitbol2017, Bryan2018, Buzzelli2018, SO2018}. This growth comes mainly from a surge in the number of detectors deployed per focal plane, which in turn is facilitated by telescope designs optimised for large fields of view (FOV) \citep{Niemack2016}. Increased sensitivity requirements also motivate extensive in-situ instrument characterisation, a time consuming process for wide FOV telescopes and satellites with limited observing time. This prompts the development of advanced modelling and analysis techniques that maximise the observing duty cycle. 

Traditionally, optical designs for CMB telescopes are optimized for high Strehl numbers, and other geometrical aspects such as $f$-number, telecentricity, and mapping speed \citep{Page2003, Ruhl2004, Fowler2007, Aikin2010, Niemack2016, Young2018}. Although these geometrical properties are definite predictors of some optical non-idealities, we argue that the CMB telescope design process should incorporate physical optics in conjunction with fast convolution techniques, and that the need for integrating this aspect into the design is growing with the cost and sensitivity requirements of future experiments. Unfortunately, modeling and computational challenges can significantly restrict telescope design iterations that incorporate full-sky beam convolution and realistic scan strategies to assess the impact of optical non-idealities on maps, power spectra, and cosmological analyses.

Convolution algorithms for realistic beams have been discussed extensively in the CMB literature. \cite{Wandelt2001} introduced an efficient method that takes advantage of fast inverse spherical harmonic transforms and sparsity of the harmonic representation of the beam; the generalization to the polarized case was presented in \cite{Challinor2000}. An implementation of this method, described in \cite{Prezeau2010}, has been used for the \emph{Planck} analysis and is closely related to the implementation discussed in this work. Parallel to these methods, algorithms that work in the pixel domain have seen use in the \emph{WMAP} and \emph{Planck} analyses and are discussed in \cite{Wehus2009} and \cite{Mitra2011}. Approaches that focus on providing an efficient convolution operator for experiments with several thousand or more detectors have been formulated in \cite{Elsner2013} and \cite{Bicep2_III}. Finally, \cite{wallis_2014} and \cite{Hivon2017} present extensions to the pseudo-$C_{\ell}$ power spectrum estimation framework that take into account the effects of beam non-idealities.

In this work, we aim to address the issue of accurately simulating optical systematic effects for current and upcoming CMB polarization experiments. 
We describe an open-source full-sky beam convolution code library that may be used to efficiently simulate time-ordered data and probe various optical systematics. We argue that although simulating time-ordered data is computationally intensive compared to the pseudo-$C_{\ell}$ extensions mentioned above, it provides a useful complementary method that is uniquely capable of quantifying optical systematic effects for analyses that do not rely solely on the angular power spectrum. 
Important examples are foreground characterisation, lensing and non-Gaussianity estimation. 
Additionally, by working in the time domain, optical effects can be simulated without having to make assumptions about other systematic effects such as non-trivial noise properties and high-level analysis choices like time-domain filtering and map-making algorithms. 
We address some of the associated numerical challenges faced by experiments that deploy a large number of detectors coupled to large-aperture optics. 
The code library is publicly available and accessible on GitHub.\footnote{\href{https://github.com/adrij/beamconv}{https://github.com/adrijd/beamconv}}

\begin{figure}
\begin{center}
\includegraphics[width=6cm]{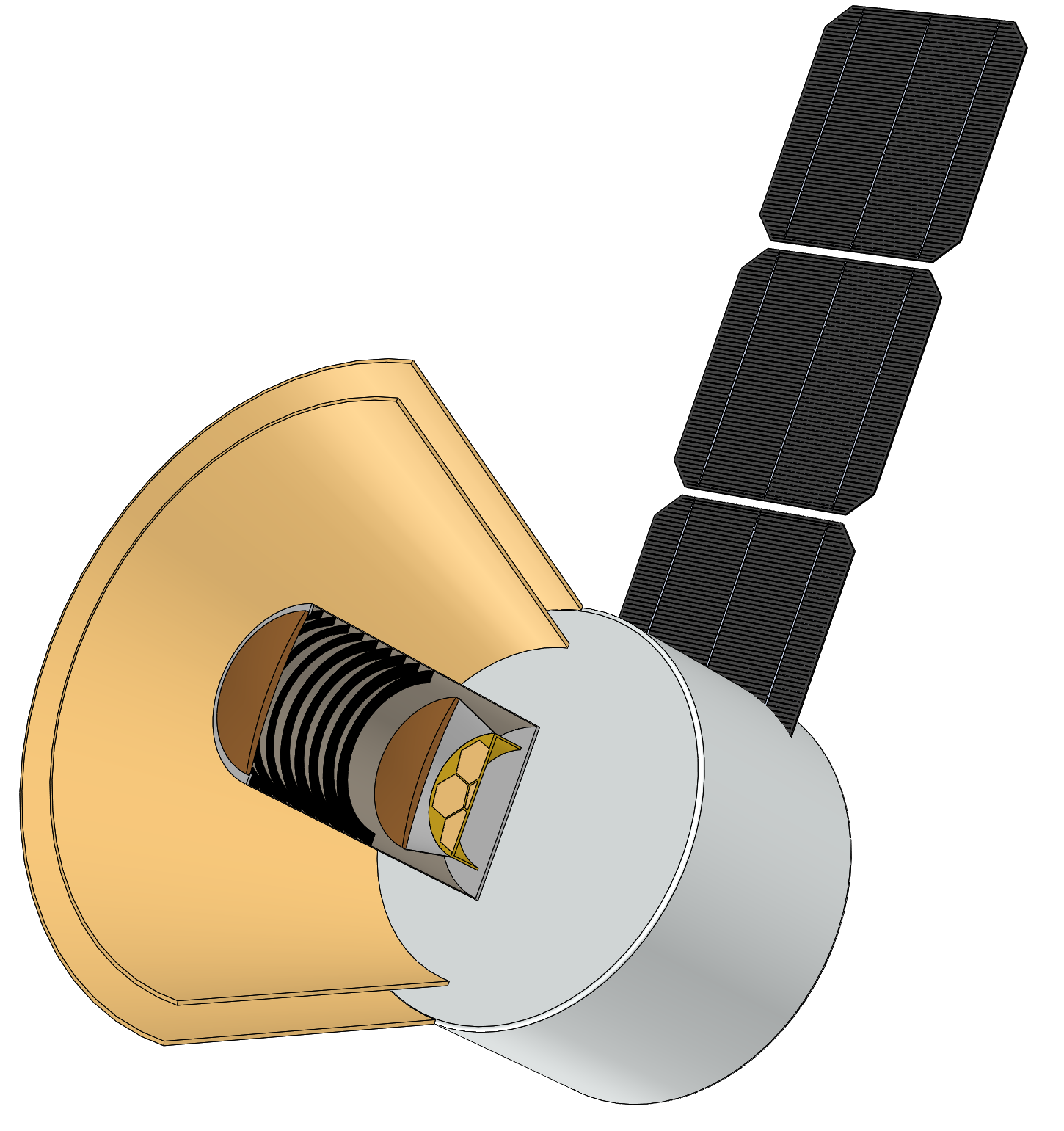}
\caption{A very rough CAD model showing the fiducial satellite design and refractive optics. The two-lens refracting telescope and sun shields are shown with a section view to emphasise some relevant components, including the two lenses (brown) and the location of the hexagonal detector tiles at the focal plane.}
\label{fig:satellite}
\end{center}
\end{figure}

We use the code library in conjunction with physical optics simulations to demonstrate its capabilities and to quantify some of the systematics faced by a fiducial satellite experiment designed to study the polarization of the CMB on degree angular scales (see Figure \ref{fig:satellite}).  We discuss the relative contributions of some of the optical systematics that are intrinsic to the proposed optical design. Although we try to identify some key questions and challenges associated with the design, this paper only covers a very small set of non-idealities formed by the interplay between detectors and refractive optics. The large number and varied properties of CMB telescope optical elements, including lenses, reflectors, baffles, filters, birefringent crystals, etc., can lead to serious modeling challenges. In fact, accurate modeling of complete optical systems is still markedly limited by computation and memory requirements as well as uncertainties in material properties. 


The paper is organized as follows. In Sec.~\ref{sec:math} and~\ref{sec:code} we introduce the mathematical formalism and present the code implementation. In Sec.~\ref{sec:instrument} the fiducial instrument is described and motivated. Results are shown in Sec.~\ref{sec:results} and, finally, discussions and suggestions for future work are presented in in Sec.~\ref{sec:conclusions}.

\section{Formalism}
\label{sec:math}


\subsection{Preliminaries}
\label{sec:math_prelims}

	We describe the polarization state of quasi-monochromatic radiation at wavelength $\omega$ originating from the far-field of a telescope by a complex vector field $\epsilon$ with a redundant overall complex phase. We represent $\epsilon$ as a vector field on the celestial sphere, i.e. we have $\epsilon(\omega) = \epsilon^i(\omega) \hat{e}_{(i)}$ with $i\in \{1,2\}$ and $ \hat{e}_{(i)}$ the basis vectors of the tangent space $T_x$ with $x\in S^2$. Most modern CMB polarization experiments use incoherent detectors, so we will restrict ourselves to this case. The incoherency of the detectors refers to their insensitivity to the phase, frequency, and polarization state of incident radiation, meaning that the vector field $\epsilon$ is not an observable. Intrinsically, these detectors are only sensitive to the total intensity of the field: $I = \langle \epsilon_i \overline{\epsilon}^i \rangle$.\footnote{We implicitly sum over repeated indices; explicit summation will sometimes be used for clarity.} For the purpose of CMB polarimetry, some polarization sensitive interface like an antenna is coupled to the detectors. This allows them to probe the cosmologically relevant quantity: the covariance of the field: $W_{ij} = \langle \epsilon_i \overline{\epsilon}_j \rangle$. The incident radiation is thus naturally described by this tensor-valued field on the sphere. By introducing an orthonormal coordinate frame, the field can be decomposed into the four (real-valued) Stokes parameters. For example: with the standard $(\theta, \phi)$ spherical coordinate system we get:
\begin{align}\label{eq:density_mat}
W_{ij}(\theta, \phi, \omega) = \frac{1}{2} \begin{pmatrix*}[l]  I+Q &  (U - iV)  \sin \theta \\ (U + iV) \sin \theta  & (I - Q) \sin^2\theta  \end{pmatrix*} (\theta, \phi, \omega) \, .
\end{align}

The $I$ and $V$ Stokes parameters represent the total intensity and circular polarized radiation component. Linear polarization is described by $Q$ and $U$.\footnote{
Throughout this work whenever we work with the spherical coordinate system or Euler angles, we will conform to the `cosmo' polarization angle convention used by the \texttt{HEALPix} library (\href{https://healpix.jpl.nasa.gov}{https://healpix.jpl.nasa.gov}) and use the ZYZ convention (with fixed axes) for Euler angles $(\psi, \theta, \phi)$. Unless noted otherwise, we use spherical coordinates with basis vectors $\hat{e}_{(\theta)} = \partial_{\theta}$, $\hat{e}_{(\phi)} = \partial_{\phi} / \sin \theta $ and metric $g_{ij} = \mathrm{diag}(1, \sin^2\theta)$.} 
It is important to realise that, because they correspond to the components of a second order tensor field, the Stokes parameters are basis dependent with transformation properties that reflect the underlying tensor transformation law. Naturally, the $I$ parameter, being the trace of $W$, remains invariant. In contrast, the $U$ and $V$ fields behave as parity-odd under reflections; the $Q$ and $U$ parameters, corresponding to the symmetric, traceless part of the tensor field, transform among themselves under rotations of the coordinate system. Because $W$ must be positive semi-definite, the Stokes parameters obey the following inequality:
\begin{align}\label{eq:stokes_crit}
I  \geq \sqrt{Q^2 + U^2 + V^2} \, ,
\end{align}
which is saturated for purely polarized light, while unpolarized light has $Q=U=V=0$.

The four Stokes parameters may be grouped into a four-vector (Stokes vector):  $s^{\mu} = \left(I, \, Q,\, U,\, V \right)$. We  define $\mathcal{S}$ as the set of valid Stokes parameters: $\mathcal{S} = \left\{S \ \Big{|}  \ I  \geq \sqrt{Q^2 + U^2 + V^2} \right\}$ and identify the linear transformations: $M: \mathcal{S} \rightarrow \mathcal{S}$, that transform valid Stokes vectors among themselves: the so-called Mueller matrices. We will later describe the instrumental effects through the use of these transformations. 
An important subset of the Mueller matrices is the set of Mueller-Jones matrices. These are transformations that could equally well be described by a $2 \times 2$ complex (Jones) operator working directly on the complex polarization state $\epsilon$. 
 Such transformations are said to be non-depolarizing, i.e. they are unable to convert a purely polarized signal to a partly polarized or unpolarized signal. 
 All other Mueller matrices describe fully or partly depolarizing transformations.\footnote{A sufficient and necessary condition for a matrix $M$ to represent a Mueller transformation is the positive semidefiniteness of the associated coherency matrix: $\bm{H} = \frac{1}{4} M^{\mu}_{\phantom{a} \nu} \left(  \bm{\sigma}_{(\mu)} \otimes  \bm{\sigma}^{(\nu)} \right)$, with $\bm{\sigma}_{\mu} = \{ \bm{1}, \bm{\sigma}_{3}, \bm{\sigma}_{1}, \bm{\sigma}_{2} \}$ in terms of the Pauli matrices. If $\bm{H}$ has just one nonzero eigenvalue, $M$ is a Mueller-Jones transformation. \citep{cloude_1986, anderson_1994}. Note that we simply use $G_{\mu \nu} = \mathrm{diag} (1,1,1,1)$ as metric.} 

In a manner similar to \cite{odea_2007}, we do not directly work with the $Q$ and $U$ Stokes parameters. We find it more convenient to work with the complex field $P \equiv Q+iU$ and its complex conjugate, as these quantities transform under the spin-weighted representations of the rotation group (see e.g.\ \citep{zaldarriaga_1997}). We will denote these alternative Stokes vectors {$p^{\mu} = ( I, P, \overline{P}, V)$}. The corresponding Mueller transformations in the space $\mathcal{P}$ of valid $p^{\mu}$ vectors are then denoted by  $\mathcal{M}: \mathcal{P} \rightarrow \mathcal{P}$.

\subsection{Data model with beam convolution}

For each of the detectors on a focal plane, the one-dimensional array of time-ordered data (TOD): \mbox{$ \bm{d}$ $ = \left\{ d_{0}, d_{1}, \dots d_{n} \right\} $} is modelled as some linear transformation $\bm{A}$ of the sky $\bm{s}$ and an additive noise component $\bm{n}$:
\begin{align}\label{eq:data_model_simple}
\bm{d} = \bm{A} \bm{s} + \bm{n} \, .
\end{align}
In the following, we will mostly ignore the noise component and focus on the transformation $\bm{A}$ by working towards a data model that includes optical effects (Eq.~\ref{beam_conv_fast}).

We start by describing the detector positioned at the instrument-side of the optical system with a Mueller transformation $\mathcal{M}$. We approximate the detector as infinitesimally small and place it at the centre of the spherical coordinate system. Without the coupling to a polarization sensitive interface, the incoherent detector is described by a perfectly depolarizing transformation, i.e.\ $\mathcal{M}^{\nu}_{ \phantom{1}{\mu}} \propto \delta^{\nu}_{0}\delta_{\mu}^{0}$. However, when the interface is included, all elements of the top row of $\mathcal{M}$ are allowed to be nonzero: $\mathcal{M}^{\nu}_{\phantom{1}{\mu}} \propto \delta^{\nu}_{0}$  \citep{Jones2007}. The data model thus becomes:
\begin{align}
d_t \propto \int_{\Delta \omega} d\omega  \, \int_{S^2} dx \, \left(\mathcal{M}_t\right)^0_{\phantom{1}{\mu}}(x,\omega) \, p^{\mu} (x,\omega)\, , \label{d_mueller}
\end{align}
where the integrals are over the frequency passband $\Delta \omega$ and the sky $S^2$. The sky signal is denoted by the (complex) Stokes vector $p$. The subscript in $d_t$ reminds us of the discrete nature of the data.\footnote{In reality, the convolution of the continuous sky signal with the finite detector time-response can lead to significant systematic effects if not taken into account (see e.g.\ \cite{planckvii_2015}).
}
We denote the nonzero elements of $\mathcal{M}$: $ \left(\mathcal{M}_t\right)^0_{\phantom{1}{\mu}} = (\widetilde{I},\, \overline{\widetilde{P}},\, \widetilde{P},\, \widetilde{V}  )$ as they transform like a complex Stokes vector $\overline{p}_{\mu}$. Together they should be interpreted as the beam of the detector. These elements obey the Mueller transformation requirement:
\begin{align}\label{eq:stokes_crit_beam}
\widetilde{I} \geq \sqrt{|\widetilde{P}|^2 + \widetilde{V}^2} \, , \quad \text{or equivalently:} \quad \widetilde{I} \geq \sqrt{\widetilde{Q}^2  + \widetilde{U}^2 + \widetilde{V}^2} \, .
\end{align}

By expanding Eq.~\ref{d_mueller}, we obtain: 
\begin{align}
\begin{split}
d_t \propto \int_{S^2} dx \, \Bigg[ \widetilde{I}_t(x) I (x) + \Re \Big(\overline{\widetilde{P}}_t(x) P (x) \Big) + \widetilde{V}_t(x) V (x) \Bigg] \, , \label{tod_fixed_p}
\end{split}
\end{align}
where we have suppressed the integral over and dependence on the wavenumber $\omega$. We now express the elements of the instrument's Mueller matrix in Eq.~\ref{tod_fixed_p} in terms of (spin-weighted) spherical harmonic (SWSH) coefficients (see Appendix~\ref{app:kernel}). 
We do the same for the Stokes parameters of the sky and make use of the orthonormality of the SWSHs to arrive at:
\begin{align}\label{harm_tod_gen}
d_t &\propto \sum_{\ell , m} \left[ \overline{b^{\widetilde{I}}_{\ell m, t}} a^{I}_{\ell m} +  \Re \Big(  \overline{{}_2 b^{\widetilde{P}}_{\ell m, t}} \, {}_{2} a^{P}_{\ell m} \Big) + \overline{b^{\widetilde{V}}_{\ell m, t}} a^{V}_{\ell m}   \right] \, .
\end{align}
We now impose that the only difference between the optical response at samples $t$ and $t'$ is the direction and orientation of the telescope with respect to the sky. Under a  generic rotation $g^{-1} \in SO(3)$ of the coordinate system, the spin-weighted harmonic coefficients transform among themselves as:
\begin{align}\label{eq_alm_rot}
{}_s f_{\ell m} \mapsto \sum_{m'=-\ell}^{\ell} {}_s f_{\ell m'} \, D^{\ell}_{m m'} (g) \, ,
\end{align}
where $D^{\ell}(g)$ are the $(2 \ell + 1) \times (2\ell +1 )$ Wigner $D$-matrices. We may thus compute the harmonic coefficients of the beam in some fiducial reference frame \mbox{--- the instrument frame ---} and transform to a coordinate system fixed on the sky using the above relation. Note that $g$ is continuously changing due to the scanning motion of the telescope. Doing so, we obtain the final expression for the beam-convolved TOD:
\begin{alignat}{2}
d_t \propto& \sum_{\ell , m , s} &&\left[ b^{\widetilde{I}}_{\ell s} a^{I}_{\ell m} +  \frac{1}{2} \left(  {}_{-2} b^{\widetilde{P}}_{\ell s} \, {}_{2} a^{P}_{\ell m}  + {}_{2}b^{\widetilde{P}}_{\ell s}\, {}_{-2} a^{P}_{\ell m} \right) + b^{\widetilde{V}}_{\ell s} a^{V}_{\ell m}   \right] \nonumber \\ 
& &&\times (-1)^m  D^{\ell}_{-m s}(g_t) \, , \label{beam_conv_wigner} \\
 =& \sum_{\ell , m , s} &&\left[ b^{\widetilde{I}}_{\ell s} a^{I}_{\ell m} +  \frac{1}{2} \left(  {}_{-2} b^{\widetilde{P}}_{\ell s} \, {}_{2} a^{P}_{\ell m}  + {}_{2}b^{\widetilde{P}}_{\ell s}\, {}_{-2} a^{P}_{\ell m} \right) + b^{\widetilde{V}}_{\ell s} a^{V}_{\ell m}   \right] \nonumber \\
& &&\times q_{\ell} \, e^{-is \psi_t} {}_sY_{\ell m} (\theta_t, \phi_t) \, , \label{beam_conv_fast}
\end{alignat}
where we have defined: 
\begin{align}\label{qell}
q_{\ell} \equiv \sqrt{\frac{4\pi}{ 2 \ell +1}} \, ,
\end{align}
and where $(\psi_t, \theta_t, \phi_t)$ are Euler angles parametrizing the rotation $g_t$. To arrive at the second line, we have used the relation between the Wigner $D$-matrices and the spin-weighted spherical harmonics in terms of Euler angles (see Eq.~\ref{D_as_sYlm}). When formulated like Eq.~\ref{beam_conv_wigner}, a useful interpretation of the TOD emerges: the expression is simply an inverse Wigner transform with harmonic coefficients given by the terms in the square brackets, implying that the TOD are just discrete samples from a scalar field $d(g_t)$ on the manifold given by the rotation group $SO(3)$ \citep{Wandelt2001}. Intuitively, the derived expression is simply a generalisation of the standard convolution theorem. The formulation in terms of Euler angles in Eq.~\ref{beam_conv_fast} allows for an efficient numerical implementation of the operation (see Sec.~\ref{sec:conv_numerical}).

\subsection{Beams}

The three fields on the sphere: $\{\widetilde{I},\, \widetilde{P},\, \widetilde{V}\}$, that describe the instrumental beam in the above discussion are allowed to be independent as long as they conform to  the constraint in Eq.~\ref{eq:stokes_crit_beam}. Of course, in a realistic case the fields are highly dependent; here we will discuss less general, but useful beam parameterizations.

\subsubsection{Co- and cross-polarized beams}\label{sec:co_cross_beams}

Polarized receivers are commonly characterised by their response to an electric field $\bm{\epsilon}_{\mathrm{co}}$ aligned to a reference direction (the co-polar response) and their response to the orthogonal field $\bm{\epsilon}_{\mathrm{cx}}$ (the cross-polar response). Clearly, co- and cross-polar responses are coordinate-dependent properties; in the case of linear polarization the co- and cross-polar basis is, by convention, the Ludwig-III basis \citep{ludwig_1973}. In terms of the standard spherical basis, the unit vectors of this frame are given by:
\begin{align}
\hat{e}_{(\mathrm{co})} &= \sin (\phi)  \, \hat{e}_{(\theta)} + \cos (\phi) \, \hat{e}_{(\phi)}  \, ,\label{e_co} \\
\hat{e}_{(\mathrm{cx})} &= \cos (\phi)  \, \hat{e}_{(\theta)} - \sin (\phi) \, \hat{e}_{(\phi)}  \, \label{e_cx}.
\end{align}
The Ludwig-III basis has just a single coordinate singularity that can be placed in opposite direction to the beam centre in the detector's frame of reference. The beam centre is then in the $\hat{z}$ direction where the coordinate system resembles a Cartesian system. 

In the case where the optical response is completely described by the co- and cross-polar response, the instrumental Mueller transformation in Eq.~\ref{d_mueller} is the top row of a Mueller-Jones transformation. 
Simulations of the optical system, like the ones described in Sec.~\ref{sec:po}, can be used to estimate the optical response in this regime (see Appendix~\ref{app:grasp2jones}).

Linear polarization instruments are generally designed to have minimal cross-polar response, and thus instrumental beams are often approximated by just the co-polar response. In this case, the response to circular polarization $\widetilde{V}$ vanishes while the polarized beam $\widetilde{P}$ is completely determined by the unpolarized beam $\widetilde{I}$ and a reference angle $\gamma$ to the co-polar direction: the polarization angle. Using the Ludig-III basis (indicated by subscript ${}_{\mathcal{L}}$), we then have: 
 \begin{align}\label{co_pol_appr}
  \widetilde{I}_{\mathcal{L}}(x) e^{\pm2i \gamma} &= \widetilde{P}_{\mathcal{L}}(x) \,  \quad \quad \text{(co-pol. approx.)} \, .
 \end{align}
Using Eq.~\ref{e_co}-\ref{e_cx}, one can show that the harmonic coefficients of the $\widetilde{P}$ beam in the $(\theta, \phi)$ basis are related to those of the  $\widetilde{I}$ beam by convolution with a harmonic kernel $\bm{K}$ \citep{Hivon2017}. When the support of the beam is small compared to the curvature of the celestial sphere, $\bm{K}$ is well approximated as diagonal per azimuthal mode $m$  (see Appendix~\ref{app:kernel}):
\begin{align}
{}_{\pm2}b^{\widetilde{P}}_{\ell m}  &= e^{\pm 2 i \gamma} \sum_{l'} b^{\widetilde{I}}_{\ell' (m\pm2)} K_{\ell \ell' m} \, , \\
&\approx  e^{\pm 2 i \gamma} b^{\widetilde{I}}_{\ell (m\pm2)} \, . \label{eq:co_pol_diag}
\end{align}

\subsubsection{Azimuthally-symmetric beams}

The main beam of a well-behaved polarimetric instrument is often well approximated as being azimuthally symmetric; the beam can be described as a function of angular distance to the beam centre only. For the harmonic modes of the \mbox{spin-$0$} $\widetilde{I}$ and $\widetilde{V}$ fields, this means that only the $m=0$ azimuthal modes are nonzero when the beam is placed on either pole of the $(\theta, \phi)$ coordinate system. The case for the spin-$2$ field $\widetilde{P}$ is less obvious due to the coordinate singularities at the poles. In  Appendix~\ref{app:az_symm} we demonstrate why the only nonzero modes of the $\widetilde{P}$ field are $m = \pm2$.
As a result, the harmonic coefficients of the Stokes parameters on the $(\theta, \phi)$ basis for an azimuthally symmetric beam centred on the pole obey:
\begin{align}
b^{\widetilde{I}/\widetilde{V}}_{\ell m}  &\propto \delta_{m0} \, ,\label{eq:unpol_sym}\\ 
{}_{\pm2}b^{\widetilde{P}}_{\ell m}  &\propto \delta_{m\mp2} \, \label{eq:pol_sym}.
\end{align}
This holds independently of approximating the beams as non-depolarizing or co-polar only.

In cases where the azimuthal symmetry is weakly broken, e.g.\ for detectors on the corners of a focal plane, or when a symmetric beam is not centred exactly on the pole due to detector pointing miscalibration, only a limited number of azimuthal ($m$) modes are usually needed to accurately describe the beam. In such cases, the data model in Eq.~\ref{beam_conv_fast} still makes use of the relative sparsity of the harmonic representation, as the sum over $s$ does not need to run over all $2 \ell_{\mathrm{max}} +1$ formally required values.

\subsubsection{Gaussian and Elliptical Gaussian beams}

At first order, the co-polar beam is generally well approximated by the diffraction pattern from a circular aperture. The centre region of the resulting Airy beam pattern is in turn shaped closely like an azimuthally symmetric Gaussian function with harmonic coefficients given by \citep{Challinor2000}:
\begin{align}
b^{\widetilde{I}}_{\ell m} = \sqrt{\frac{2 \ell +1}{4 \pi}} \exp \left[- \frac{\ell (\ell +1 ) \sigma^2}{2} \right] \delta_{m0} \, .
\end{align}
In the same vein, the main beam of a detector placed far off-axis on a focal plane could be approximated by an elliptical Gaussian. Closed form expressions for the corresponding harmonic coefficients can be found in \cite{Souradeep2001} and \cite{mitra_2004}.

\subsubsection{Ghosting response}
\label{sec:ghosting}
Internal reflections in a receiver, for example between lenses and focal plane, can create so-called ghost beams; mirror images of the main beam rotated away from the main beam centre \citep{Fowler2007, Aikin2010}. Optical ghosting is partially worrisome for high index-of-refraction materials such as silicon, necessitating advanced anti-reflective (AR) coating and detailed modelling and characterisation programs. Some CMB experiments using refracting telescopes have developed simulations to probe systematics caused by this effect \citep{MacTavish2008, Bicep2_III}. 
The ghost contribution can simply be added to the fields describing the main beam and used in Eq.~\ref{beam_conv_fast}. 
While this method is conceptually convenient, the large number of azimuthal modes required to accurately describe the resulting azimuthally asymmetric beam make it numerically inefficient. An alternative approach wherein the ghost beam is effectively treated as a separate detector with its own pointing coordinates is therefore generally more efficient.

\subsection{Modulation techniques}

To reduce their dependence on accurate instrumental characterisation, current and future CMB experiments often incorporate modulation techniques that reduce the degeneracies between spurious systematic signal and the sky signal. We briefly discuss how to incorporate two common techniques: boresight rotation and half-wave-plate modulation, into the data model in Eq.~\ref{beam_conv_fast}.

\subsubsection{Boresight rotation}

Boresight rotation refers to physically rotating the telescope (stepwise) around the optical axis, or boresight. Having access to redundant observations made at different boresight angles is beneficial in many aspects. See e.g.\ the BICEP experiment and its successors \citep{bicep_2010, bicepkeck_2015, bicep3_2016}. In terms of optical systematics, its main purpose is to suppress temperature-to-polarization leakage due to azimuthally asymmetric modes of the $\widetilde{I}$ beam (see Sec.~\ref{mapmaking}). Boresight rotation is most naturally included in the data model by including it in the pointing: $(\psi_t, \theta_t, \phi_t)$ (in Eq.~\ref{beam_conv_fast}), while leaving the beam coefficients unchanged.

\subsubsection{Half-wave plate modulation}\label{sec:hwp_mod}

A half-wave plate (HWP) is a birefringent material that changes the polarization state of incoming radiation of a specific wavelength and incidence angle by introducing a phase difference of $\pi$ between the radiation component aligned along a direction intrinsic to the material (the fast axis) and the orthogonal component. Notably, for incident linearly polarized light, the effect is to mix $Q$ and $U$ by an amount based on the orientation of the HWP's fast axis in reference to the coordinate frame defining the Stokes parameters (see Sec.~\ref{mapmaking}). Rotating the HWP thus results in a controlled modulation of the incoming linear polarization.

Half-wave plate modulation in the context of CMB polarimeters has been discussed in e.g.\ \cite{odea_2007}, \cite{MacTavish2008} and \cite{brown_2009}. 
In terms of suppressing optical systematics it differs qualitatively from the boresight rotation discussed in the above; both techniques effectively result in a controlled modulation of the linearly polarized signal of the sky, but boresight rotation does not leave the intensity signal unchanged in the case of azimuthally asymmetric beams.
In contrast, an (ideal) HWP placed skywards of the telescope will leave the signal induced by the intensity beam unchanged, regardless of its shape, thus decoupling it from the modulated linearly polarized sky signal. 

At subleading order, non-idealities in the HWP will spoil this behaviour by making the $\widetilde{I}$ and $\widetilde{V}$ beams weakly dependent on the HWP angle. See e.g.\ \citep{savini_2006, bryan_2010, essinger-hileman_2016}. In terms of the data model in Eq.~\ref{beam_conv_fast}, HWP modulation is thus most generally described by beam coefficients that depend on the HWP angle. In case of an ideal skywards HWP however, the coefficients may be factored into two terms: one that does and one that does not depend on the HWP modulation angle:
\begin{align}
{}_{\pm2}b^{\widetilde{P}}_{\ell m} \rightarrow {}_{\pm2}b^{\widetilde{P}}_{\ell m} e^{\pm 4 i \alpha} \quad \quad (\mathrm{ideal \: HWP \:  modulation}) \, , \label{eq:ideal_hwp_blm}
\end{align}
where $\alpha$ is the HWP angle. 
The $\widetilde{V}$ coefficients simply pick up a minus sign ($b^{\widetilde{V}}_{\ell m} \rightarrow - b^{\widetilde{V}}_{\ell m}$) when the HWP is introduced and the $\widetilde{I}$ coefficients remain unchanged.

\subsection{Systematics arising from map-making}\label{mapmaking}

After data acquisition, the polarized sky signal is reconstructed by solving the inverse problem associated with Eq.~\ref{eq:data_model_simple}. Generally this is done by calculating a point estimate of the sky signal (a pixelized map) in a process called map-making. Commonly, the map-making estimator is a variation on the generalized least squares statistic, given by:
\begin{align}\label{eq:least_squares}
\bm{\hat{s}} = \left( \bm{A}^{\dagger} \bm{N}^{-1} \bm{A} \right)^{-1} \bm{A}^{\dagger} \bm{N}^{-1} \bm{d} \, .
\end{align}
In case of Gaussian noise with an a priori known covariance $\bm{N}$ and uniform signal prior in the specified basis, this corresponds to the maximum a posteriori estimate. The (Gaussian) posterior around the maximum is then described by the $ \left( \bm{A}^{\dagger} \bm{N}^{-1} \bm{A} \right)^{-1}$ covariance matrix. In realistic analyses, this matrix is too large and dense to be available for regular matrix calculations, but its operation on a map-sized array (as in Eq.~\ref{eq:least_squares}) can be calculated iteratively.\footnote{
This model can be extended by jointly inferring the noise covariance \citep{prunet_2001, natoli_2002, wehus_2012} or signal covariance \citep{wandelt_2004, eriksen_2008, taylor_2008}. Including statistical inference on part of the transformation $\bm{A}$ (e.g. the beam) in such approaches is relatively unexplored. 
}

Reconstruction of the three- or four-dimensional signal Stokes vector\footnote{
The fourth Stokes parameter $V$ is often ignored in CMB data analysis as the cosmological signal is not expected to be significantly circularly polarized \citep{king_2016} and instruments are designed to be insensitive to it. Still, in the context of optical systematics, $V$ cannot be entirely ignored. For instance, Zeeman splitting of oxygen in the Earth's magnetic field at $60$ and $118.8$ $\mathrm{GHz}$ provides a significant source of circular polarization for ground-based experiments \citep{hanany_2003,hanany_2013} which could be converted to linear polarization by non-ideal half-wave plates or other significant cross-polar responses~\citep{nagy_2017}.}
from the one-dimensional data crucially relies on knowledge of the linear transformation ($\bm{A}$) that maps the sky signal onto the time domain. Any beam related systematics will come from an incorrectly assumed transformation. In the above, we have shown how, besides telescope pointing, calibration, sample flagging and other instrumental effects, the transformation should contain the beam-convolution. However, this aspect is often ignored in the map-making stage because the a priori knowledge of the beams is too poor to include them in a point estimate $\hat{\bm{s}}$. Furthermore, for azimuthally symmetric beams, it is simpler to forward-propagate the effects by convolving a model of interest (sky map, power spectrum, etc.) with the beam.  

\begin{figure*}
\begin{center}
\includegraphics[width=17cm]{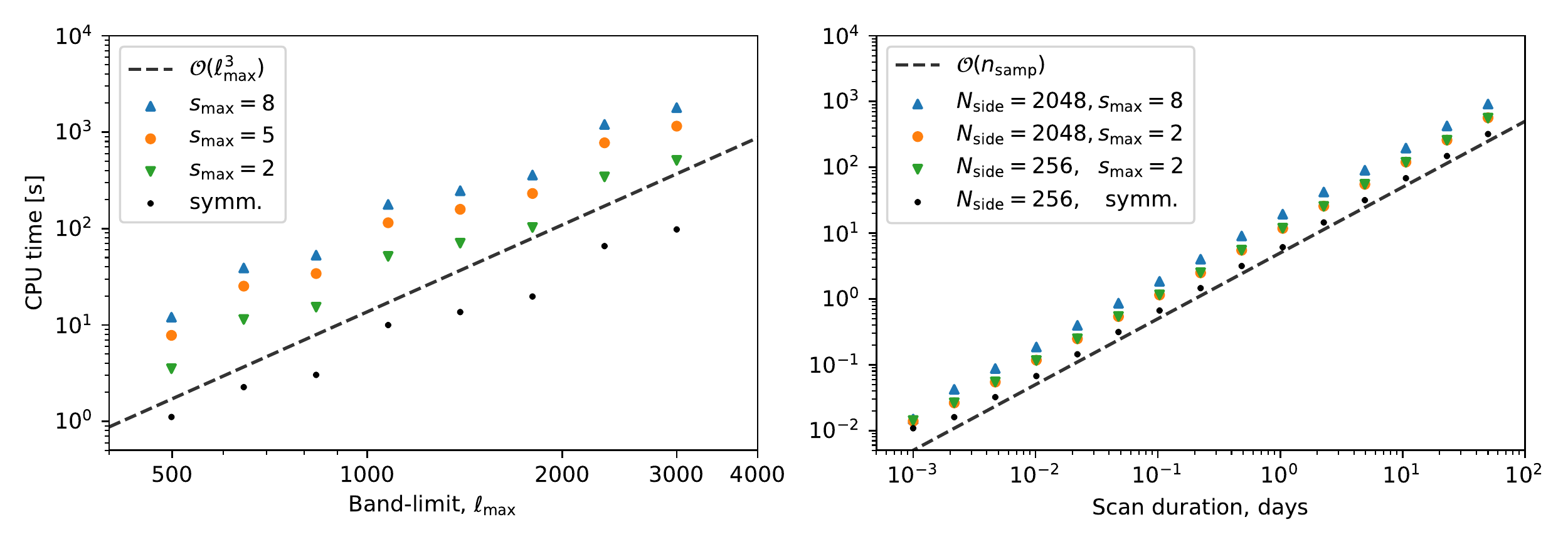}
\caption{Left: log-log plot of required CPU time for the inverse spin-weighed spherical harmonic transforms needed for a single (linearly-polarized) beam as function of band-limit $\ell_{\mathrm{max}}$. 
The different marker types refer to the azimuthal band-limit ($s_{\mathrm{max}}$) of the beam. The black dots correspond to convolution with an azimuthally symmetric beam.
The dashed line shows the expected asymptotic scaling (with arbitrary normalisation). 
The results conform relatively well with the expected scaling, but show small, step-like deviations due to changes in the pixelisation scheme (we let $N_{\mathrm{side}}$ be the smallest power of $2$ that is larger than $\ell_{\mathrm{max}} / 2$). 
Results are from a single thread on an Intel Xeon E5-2697 v2 core running at 2.7~GHz.
Right: log-log plot of required CPU time for producing time-ordered data as function of the scan duration (on the same single core setup). 
We use a sampling frequency of $100$~Hz with pointing quaternions and beam-convolved maps preloaded in memory. 
The scaling follows the expected linear relation with the number of time-samples $n_{\mathrm{samp}}$ (illustrated by the arbitrarily normalised dashed line). 
The required CPU time is largely independent of the number of pixels ($n_{\mathrm{pix}} = 12N_{\mathrm{side}}^2$) of the convolved maps and weakly linearly dependent on the azimuthal band-limit $s_{\mathrm{max}}$ of the beam. Again, the black dots denote the azimuthally symmetric case.
No interpolation is used while sampling the data.}
\label{fig:computation_time}
\end{center}
\end{figure*}

We can gain some intuition for the effects of different beam non-idealities by considering limiting cases of the map-making procedure. 
We start by  approximating the noise covariance as diagonal in the time sample domain (white noise): $\bm{N} = \langle n_t n_t' \rangle \propto \delta_{t t'}$. Secondly, we describe the problem in the space $\mathcal{P}$ of complex Stokes vectors $p^{\mu}$ (see Sec.~\ref{sec:math_prelims}). The map-making estimate then reduces to: $\bm{\hat{p}} \propto \left( \bm{A}^{\dagger} \bm{A} \right)^{-1} \bm{A}^{\dagger} \bm{d}$ per pixel on the sky; we will refer to this as binning map-making. An ansatz for $\bm{A}$ that ignores the beam but is otherwise valid for a single co-polarized detector with vanishing polarization angle can be derived from \cite{Jones2007}:
\begin{align} \label{eq:naive_mapmaking}
\bm{A} = \left(A_{t,x}\right)^{\mu} \propto \left(1, \frac{1}{2}e^{-2i\psi_{t}}, \frac{1}{2}e^{+2i\psi_{t}}, 0\right) \bm{1}_{X}(t) \, .
\end{align}
The position angle $\psi_t$ is included in Eq.~\ref{beam_conv_fast} and the indicator function $\bm{1}_{X}(t)$ is defined to be $1$ for time samples $t$ in the set $X$ of samples that hit pixel $x$ and zero otherwise. We will restrict ourselves to the estimate $\hat{p}^1 = \hat{P}_x$: the linearly polarized signal in pixel $x$.
\begin{align}
\hat{P}_x  \propto  [ ( \bm{A}^{\dagger} \bm{A} )^{-1}]^{1}_{\phantom{1}\nu} \sum_{t\in P}  (A_{t, x})^{\nu} d_{t} \, .  \nonumber
\end{align}
The normalisation is then given by the inverse of:
\begin{align}
[  \bm{A}^{\dagger} \bm{A} ]^{\mu}_{\phantom{1}\nu} &\propto  \frac{1}{2} \sum_{t\in X} \begin{pmatrix} 
2 &  e^{-2i\psi_{t}} &  e^{+2i\psi_{t}} & 0 \\
e^{+2i\psi_{t}} & \frac{1}{2} & \frac{1}{2}e^{+4i\psi_{t}} & 0 \\
e^{-2i\psi_{t}} & \frac{1}{2}e^{-4i\psi_{t}} & \frac{1}{2} & 0 \\
0 & 0 & 0 & 0
\end{pmatrix} \, .  
\end{align}
We  focus on a scan strategy that visits the pixel with a large uniformly distributed set of position angles $\psi$. Summing over $t$ then diagonalizes the above, which results in a diagonal inverse matrix after we project out the singular $V$ part by taking the pseudoinverse. After inserting the expression for $d_t$ (Eq.~\ref{beam_conv_fast}), the estimate for $P$ in pixel $x$ becomes solely proportional to the $m=2$ modes of the beam:
\begin{align}
\begin{split}
\hat{P}_x \propto \frac{1}{2} \sum_{\ell , m } \bigg\{ & b^{\widetilde{I}}_{\ell 2} a^{I}_{\ell m} +  \frac{1}{2}\left(  {}_{-2} b^{\widetilde{P}}_{\ell 2} \, {}_{2} a^{P}_{\ell m}  + {}_{2}b^{\widetilde{P}}_{\ell 2}\, {}_{-2} a^{P}_{\ell m}  \right) \\ 
&+ b^{\widetilde{V}}_{\ell 2} a^{V}_{\ell m}\bigg\} \, q_{\ell} \, {}_{2}Y_{\ell m}\Big|_x  \, .
\end{split}
\end{align}
Here $|_x$ indicates evaluation at the $\theta, \phi$ coordinates of the pixel centre and $q_{\ell}$ is given by Eq.~\ref{qell}. This expression makes explicit how a nonzero quadrupole ($m=2$) mode of the unpolarized beam biases the $\hat{P}$ estimate by modulating the dominant unpolarized sky signal just like the linearly polarized signal. 
As we are already in the limit of perfectly uniform position angle coverage, it is clear that boresight rotation cannot modulate this bias away.
Additionally, in the limit of uniform sky coverage, one can show \citep{hu_2003, odea_2007} that the real part of $b^{\widetilde{I}}_{\ell 2}$ purely sources temperature-to-$E$-mode ($I \rightarrow E$) leakage, while the imaginary part is responsible for $I \rightarrow B$ leakage.\footnote{The $E$- and $B$-mode harmonic coefficients are given by:
\begin{align}\label{eq:e_b_modes}
a_{E, \ell m} = -\frac{1}{2} ({}_{2} a^{P}_{\ell m} + {}_{-2} a^{P}_{\ell m}) \, , \quad a_{B, \ell m} = \frac{i}{2} ({}_{2} a^{P}_{\ell m} - {}_{-2} a^{P}_{\ell m}) \, .
\end{align}
In terms of covariant derivatives of the symmetric traceless (ST) part of $W$ (Eq.~\ref{eq:density_mat}) we have $a_{E,\ell m} = N_{\ell} \int dx \nabla^{i} \nabla^{j} (W_{\mathrm{ST}})_{i j} \overline{Y}_{\ell m}$ and $a_{B,\ell m} = N_{\ell} \int   dx \,\widetilde{\epsilon}^{i}_{\phantom{a}k} \nabla^{k} \nabla^{j} (W_{\mathrm{ST}})_{ij} \overline{Y}_{\ell m}$ ($\widetilde{\epsilon}$ is the Levi-Civita symbol and $N_{\ell} \! \equiv \! \sqrt{2(\ell - 2)!/(\ell -2)!}$) \citep{kamionkowski_1997}, illustrating that $a_{E/B, \ell m}$ are harmonic modes of a rotational scalar and pseudoscalar field.}
In more practical terms, the real and imaginary parts of  $b^{\widetilde{I}}_{\ell 2}$ correspond to the components of the unpolarized beam with azimuthal parts proportional to $\cos 2 \phi$ and $\sin 2 \phi$ respectively.   
Only in this limit (perfect uniform coverage in $\theta$, $\phi$ and $\psi$), the leakage may be described independently per mode; leakage from $\ell$ to $\ell'$ and $m$ to $m'$ will occur in more general cases \citep{hu_2003, Hanson2010, Hivon2017}.
In the case of azimuthally symmetric beams (\mbox{Eq.~\ref{eq:unpol_sym}-\ref{eq:pol_sym}}), the estimate reduces to a symmetrically smoothed version of the signal that is independent from the $I$ and $V$ sky.
 
The situation changes when, instead of relying on rotating the instrument (or Earth's rotation), the angular information needed to solve for $p^{\mu}$ is obtained by half-wave plate modulation. In this case, the data model assumed for map-making would be expanded as follows:
\begin{align}
\bm{A} = \left(A_{t,x}\right)^{\mu} \propto \left(1, \frac{1}{2} e^{-2i(\psi_t+2\alpha_{t})}, \frac{1}{2} e^{+2i(\psi_t + 2\alpha_{t})}, 0\right) \bm{1}_{X}(t) \, ,
\end{align}
where $\alpha$ denotes the HWP angle (see Sec.~\ref{sec:hwp_mod}).
The estimate for the linearly polarized component (at fixed position angle $\psi$) then becomes proportional to:
\begin{align}
\begin{split}
\hat{P}_x \propto  \frac{1}{2} \sum_{t\in X}  \sum_{\ell , m } \bigg\{ & b^{\widetilde{I}}_{\ell s} a^{I}_{\ell m} +  \frac{1}{2}\left(   {}_{-2} b^{\widetilde{P}}_{\ell s} \, {}_{2} a^{P}_{\ell m}  + {}_{2}b^{\widetilde{P}}_{\ell s}\, {}_{-2} a^{P}_{\ell m}    \right) 
\\ &+ b^{\widetilde{V}}_{\ell s} a^{V}_{\ell m}\bigg\} \, q_{\ell} e^{4i\alpha_t}\, {}_{s}Y_{\ell m}\Big|_x  \, .
\end{split}
\end{align}
If the set of HWP angles is large and uniformm the only non-vanishing terms are proportional to $\exp(- 4i \alpha_t)$. With an ideal (skyward) HWP that is true for ${}_{-2} b^{\widetilde{P}}_{\ell s}$ (see Eq.~\ref{eq:ideal_hwp_blm}), while the $\widetilde{I}$, $\widetilde{V}$ coefficients remain constant with $\alpha$. The $\hat{P}$ estimate is then only sourced by the linearly polarized sky (regardless of the shape of the beam). Of course, subleading $E \leftrightarrow B$ leakage due to azimuthal asymmetry ($m \neq \pm 2$ modes) of the linearly polarized beam is not suppressed by HWP modulation, but requires sky or boresight rotation to be suppressed. Finally, any cross-polar components (e.g.\ a miscalibrated polarization angle) of the linearly polarized beam are not suppressed by HWP modulation, nor can its azimuthally-symmetric part be suppressed by a uniform sampling of position angle $\psi$. 


The above examples provide intuition for the cause of some of the leading order optical systematic effects. 
Another leading order effect is a simple miscalibration of the (dominant) azimuthally symmetric co-polarized part of the beam, e.g.\ by incorrectly assuming it to be Gaussian. 
Such a mistake will, on its own, not mix $I$ and $P$ or $E$ and $B$ but will still result in a wrongly inferred amplitude of anisotropies. 
This is especially problematic at small angular scales where deviations in the amplitude of the CMB power spectra are highly degenerate with varying effective beam size. 

Any realistic map-making algorithm is capable of jointly solving for the signal estimate $\bm{\hat{p}}$ using data from multiple detectors. 
Additionally, more sophisticated algorithms than those used in the examples above exist.
One common choice is the so-called pair differencing method \citep{Jones2007}. 
Here, the linearly polarized signal is directly estimated from the differenced TOD from detector pairs that share a physical location on the focal plane but are coupled to orthogonal linearly polarizing interfaces. 
The resulting estimate uses suboptimal noise weighting compared to using both detectors independently. 
However, the cancelation of common modes in the noise or unpolarized signal that may otherwise be difficult to explicitly model is advantageous. The estimate is similarly uninfluenced by common features in the $\widetilde{I}$ beams, e.g.\ a shared azimuthally asymmetric component. One can check that this also holds true when the map-maker from Eq.~\ref{eq:naive_mapmaking} is used with these paired detectors.
On the other hand, any $\widetilde{I}$ beam component that does not cancel exactly, regardless of its azimuthal dependence or spin, directly biases the $\hat{P}$ estimate by $I \rightarrow P$ leakage. 
This includes miscalibrated gain or beamwidth differences between two paired detectors \citep{Bicep2_III}. 
These two systematic effects do not result in $I \rightarrow P$ leakage when a map-making scheme like  Eq.~\ref{eq:naive_mapmaking} is used.

Finally, two other map-making approaches that attempt to correct for beam effects are worth mentioning. The first, as proposed in \cite{Bock2009} and \cite{wallis_2014}, uses an ansatz for $\bm{A}$ that is similar to Eq.~\ref{eq:naive_mapmaking}, but contains a number of additional harmonics such as $\exp (\pm i \psi_t)$ or $\exp (\pm 3 i \psi)$. The resulting map-making estimate has a higher dimension than the standard $\{\hat{I}, \hat{P} \}$ estimate and therefore projects out modes that are necessarily spurious. Of course, the method results in increased uncertainty in the $\{\hat{I}, \hat{P} \}$ estimate and is unable to project out the most problematic spurious signal: the one proportional to $\exp( \pm 2 i \psi_t)$. Another method, described in e.g.\ \cite{Armitage2009} and \cite{keihanen_2012}, imposes a maximally informative prior on the beam by directly using the full beam-convolved data model from Eq.~\ref{beam_conv_fast} as ansatz for $\bm{A}$. Computing the point estimate by solving Eq.~\ref{eq:least_squares} becomes much more involved but can still be done using the conjugate gradient method and by regularising the singular part of $\bm{A}$. Still, the method has not been demonstrated to work with non-white noise or high resolution data ($\ell_{\mathrm{max}} > 2000$). Arguably, a more significant challenge associated with this method is the one alluded to in the introduction to this section: any prior uncertainty on the beam is lost in the map-making procedure. This, together with the high numerical demands, and dependence on map-making schemes, may suggest that methods relying on forward-propagating beam effects are generally more useful than those that deconvolve the beam. 

\begin{table}
\caption{Optical properties of the two-lens silicon designed considered in this analysis. Note that $c$ and $k$ represent the inverse radius of curvature and the conic constant, respectively. The silicon lenses are assumed to have an index of refraction of $n_\mrm{si} = 3.42$ and the physical separation between primary and secondary lens is 550 mm. The focal plane is located 230 mm behind the secondary lens.\label{tab:opt_prop}}

\begin{tabular}{lllll}
\hline
Lens & surface & $c$ [m$^{-1}$] & $k$ & $f$-number \\
\hline
\multirow{2}{*}{\textbf{Primary}} & Sky & 1.446 & 0.141 & \multirow{2}{*}{1.56} \\
 & Focal plane & 0.933 & 1.369 & \\
\multirow{2}{*}{\textbf{Secondary}} & Sky & 1.635 & -0.052 & \multirow{2}{*}{1.56} \\
 & Focal plane & 0.834 & 14.841 & \\
\hline
\end{tabular}
\end{table}

\begin{figure}
\begin{center}
\includegraphics[width=8cm]{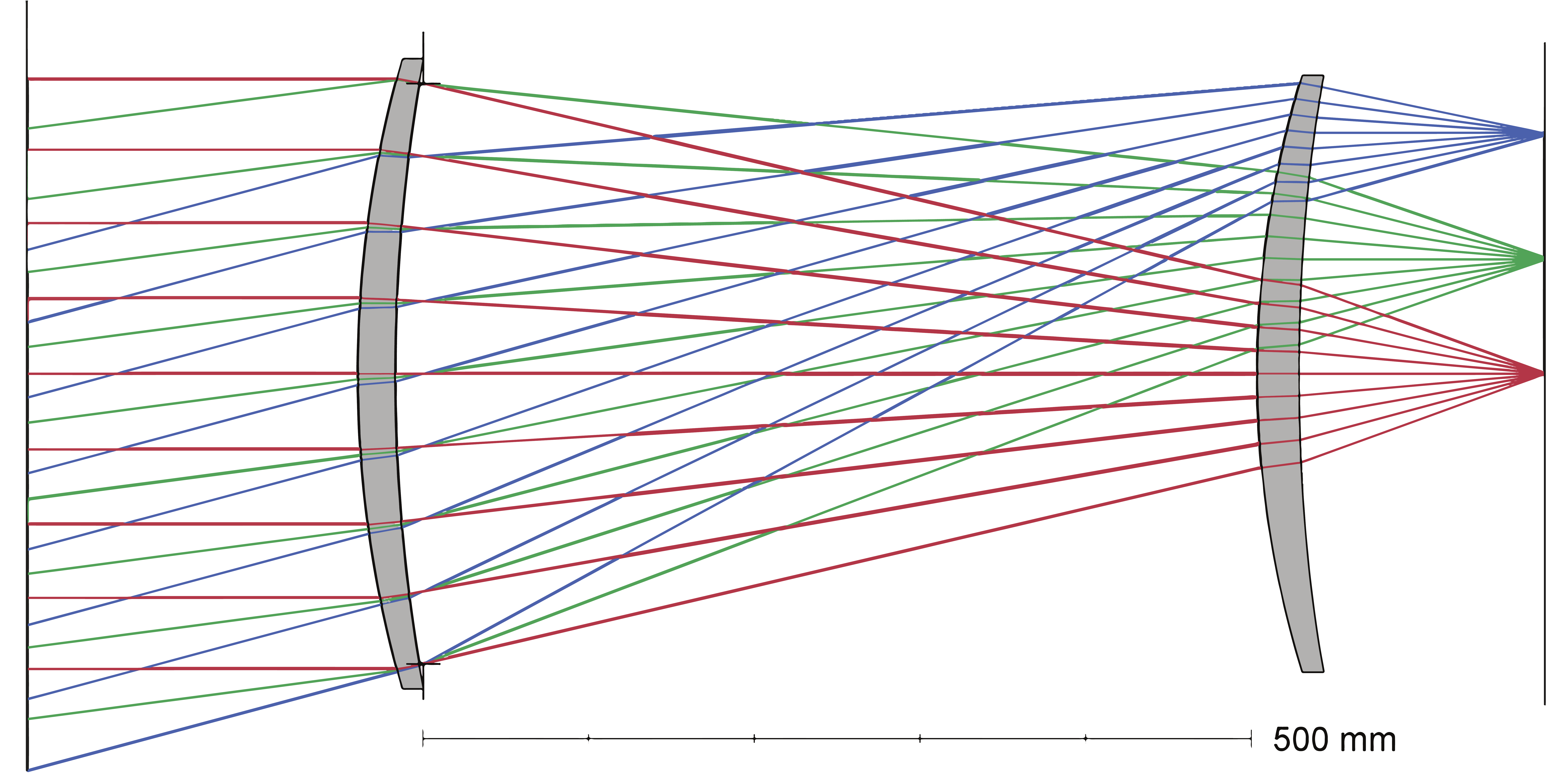}
\caption{Ray trace diagram of the proposed design showing the path of a few fields through a pair of convex-concave lenses. An optical stop is located right after the primary lens on the image (right) side. The outermost pixel (blue rays) is at a $15^{\circ}$ angle relative to boresight.}
\label{fig:ray_trace}
\end{center}
\end{figure}

\begin{figure*}
\begin{center}
\includegraphics[width=17cm]{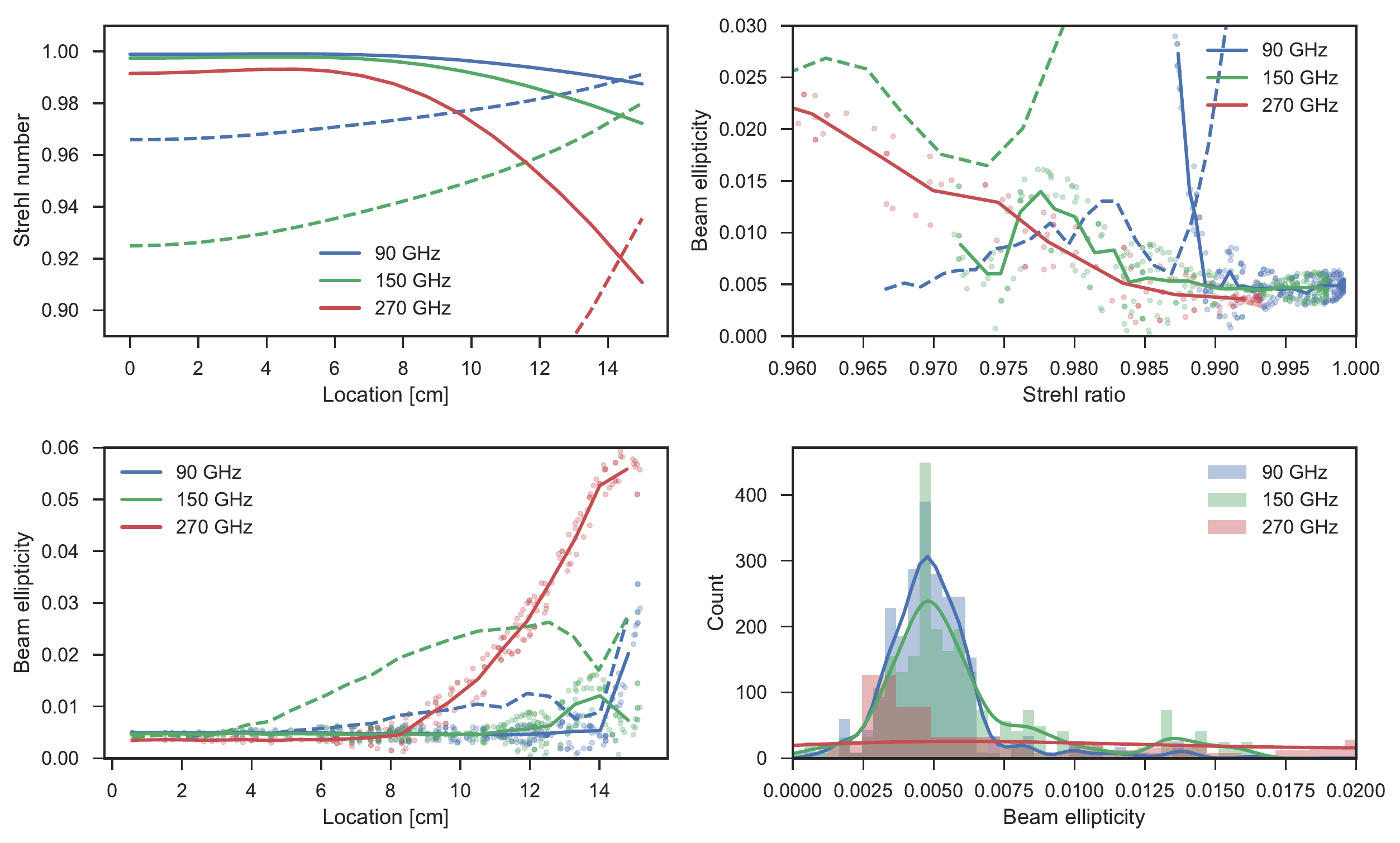}
\caption{Top left: Strehl ratio (as output by Zemax) of the proposed optical system as a function of field location for three frequencies, 90, 150, and 270 GHz, represented by top, middle, and bottom solid lines respectively. The three dashed lines correspond to the Strehl ratios for the same optical design if the silicon effective index of refraction were $\tilde{n}_{\mathrm{si}} = 3.39$ instead of $n_{\mathrm{si}} = 3.42$. Top right: Beam ellipticity as a function of Strehl ratio. Solid lines correspond to average within a fixed interval whereas the dashed lines (90 and 150 GHz) are the corresponding results for $\tilde{n}_{\mathrm{si}} = 3.39$. Lightly-coloured markers indicate the results of individual simulations (300 markers for each frequency). The markers are omitted for the case of $\tilde{n}_{\mathrm{si}} = 3.39$. Also note that we have zoomed in on the area of highest Strehl ratio which contains most of the 90- and 150-GHz detectors. It is interesting to see that the rise in ellipticity at 90 GHz appears to be dominated by diffraction effects and not a significant drop in Strehl ratio. For 150 GHz, we note a more complex behavior.
Bottom left: Beam ellipticity as s function of field location. Again, the solid line traces the average and the dashed lines correspond to the case where $\tilde{n}_{\mathrm{si}} = 3.39$. Markers identify individual simulation results. Bottom right: Distribution of beam ellipticity for all pixels simulated. At 270 GHz, the beam ellipticity distribution is relatively flat from 0-0.06 and we choose to zoom in on the distributions for lower ellipticities.}
\label{fig:strehl}
\end{center}
\end{figure*}

\section{Code description} \label{sec:code}

The primary functionality of the \texttt{beamconv} code library is to compute time-ordered data (TOD) that includes spurious signal due to optical systematics. 
The resulting TOD may be used as input to pipelines that describe further stages of data acquisition and analysis (e.g.\ addition of detector noise, time stream filtering and map-making), as beam convolution is a natural first step in any simulation pipeline.
Alternatively, the library provides simple map-making functionalities to help assess the systematic signal in the noiseless limit.

The code is written in \texttt{Python} and relies heavily on the standard scientific computing package \texttt{numpy}. 
The (inverse) spherical harmonic transforms are handled by the highly optimised \texttt{libsharp} library \citep{reinecke_2013} using an interface provided by the \texttt{healpy} \texttt{Python} package.\footnote{\href{http://healpix.sourceforge.net}{http://healpix.sourceforge.net}} 
All pointing related computations are done by interfacing with the \texttt{qpoint} library.\footnote{\href{https://github.com/arahlin/qpoint}{https://github.com/arahlin/qpoint}} 
The code is setup for parallel computing on massive distributed memory systems using the \texttt{MPI} standard.
The library is bundled with several explanatory \texttt{IPython} notebooks.

The capabilities of the library partially  overlap with those of the Time-Ordered Astrophysics Scalable Tools (\texttt{TOAST}) package.\footnote{\href{http://github.com/hpc4cmb/toast}{http://github.com/hpc4cmb/toast}}
The public version of \texttt{TOAST} has recently been upgraded with an interface to the beam convolution library \texttt{conviqt} (see \citep{Prezeau2010}) and thus should be able to produce similar results as \texttt{beamconv}.
Clearly, \texttt{TOAST} is a more extensive simulation package, that is also capable of reproducing  instrumental effects that are not optics-related. 
Instead of trying to reproduce the \texttt{TOAST} library, we aim to have \texttt{beamconv} purely focused on optical systematics and hope to provide an accessible tool that can be extended to include more optical systematic effects with relative ease.

In the following sections we will briefly go over the technical details of the convolution operation, explain the input and output of the code and provide a few benchmark results. Finally, we comment on possible future additions to the code.

\subsection{Implementation}\label{sec:conv_numerical}

The beam convolution operation is performed over the full sky as point-wise multiplication in the harmonic domain, using the expression for the data in Eq.~\ref{beam_conv_fast}. The method is thus heavily inspired by the work of \cite{Wandelt2001} and is implemented similarly to the \texttt{totalconvolver} and \texttt{conviqt} implementations of this method described in \cite{reinecke_2006} and \cite{Prezeau2010} respectively. 
We will briefly discuss our implementation. 

Calculating beam-convolved data by evaluating Eq.~\ref{beam_conv_fast} at each time sample is equally inefficient as evaluating an integral over the sphere at each sample (see Eq.~\ref{tod_fixed_p}). 
The first expression is only efficient because it allows separate treatment of the convolution and data sampling. 
To do so means that one, for each azimuthal mode $s$ of the beam, first evaluates the following inverse SWSH transformation over the entire sphere using available $\mathcal{O}(\ell_{\mathrm{max}}^3)$ algorithms:
\begin{align}\label{eq:f_s}
\begin{split}
 {}^{(s)}f(\theta, \phi) = \sum_{\ell , m } {}_{s}f_{\ell m} q_{\ell}\, {}_sY_{\ell m} (\theta, \phi) \, , \quad \forall \, \theta, \phi \in S^2 \, ,
\end{split}
\end{align}
with harmonic modes given by Eq.~\ref{qell} and:
\begin{align*}
{}_{s}f_{\ell m} =b^{\widetilde{I}}_{\ell s} a^{I}_{\ell m} +  \frac{1}{2} \left(  {}_{-2} b^{\widetilde{P}}_{\ell s} \, {}_{2} a^{P}_{\ell m}  + {}_{2}b^{\widetilde{P}}_{\ell s}\, {}_{-2} a^{P}_{\ell m} \right) + b^{\widetilde{V}}_{\ell s} a^{V}_{\ell m}   \, .
\end{align*}
Once the ${}^{(s)}f(\theta, \phi)$ maps are computed for each $s$, the TOD may be sampled from them using the $(\theta_t, \phi_t)$ pointing information and time-dependent phase given by the factor $\exp ( - i s \psi_t)$. As long as the synthesised maps can be stored in memory, data from any sort of scan strategy may be obtained. The overhead given by the inverse SWSH transforms is constant.

Given that diffraction naturally truncates the beam coefficients at some finite $\ell_{\mathrm{max}}$, the transforms only need to be computed up to $\ell_{\mathrm{max}}$. 
This can be done with an asymptotic $\mathcal{O}(\ell_{\mathrm{max}}^3)$ scaling, which will dominate the total scaling for simulation runs with large $\ell_{\mathrm{max}}$ and few data samples (see Sec.~\ref{sec:benchmarks})

Note that in \texttt{beamconv} we perform separate inverse transforms for the $\widetilde{I}$ and $\widetilde{P}$ beams (ignoring $\widetilde{V})$. This is done such that that we may modulate the linearly polarized signal independently from the total intensity component. This is, for instance, used to to incorporate time-dependent HWP modulation (see Eq.~\ref{eq:ideal_hwp_blm}).

By default, the TOD are directly sampled from (equal area) \texttt{HEALPix} pixels. We have found that this approach suffices (as long as the $N_{\mathrm{side}}$ parameter is larger than $\ell_{\mathrm{max}} / 2$)\footnote{The $N_{\mathrm{side}}$ parameter is a power of $2$ that determines the number of pixels within the \texttt{HEALPix} pixelisation scheme ($n_{\mathrm{pix}} = 12 \, N_{\mathrm{side}}^2$).}, but if needed, e.g.\ when high accuracy is needed at scales close to $ \ell_{\mathrm{max}}$, the data may be interpolated using bi-linear interpolation.
The inverse spherical transforms provided by \texttt{healpy} synthesise the harmonic coefficients onto the full sky, which is wasteful for experiments that observe small patches of the sky, but we allow this small hit in efficiency and defer an improvement to future work.\footnote{Unlike the original implementation suggested in \cite{Wandelt2001} that uses fast Fourier transforms (FFTs) in the $\theta$ and $\phi$ (and $\psi$) directions (by restating the problem on the $3$-torus instead of the $SO(3)$ rotation group), \texttt{libsharp} does not use an FFT over the $\theta$ direction. This allows one to skip latitude rings that are not visited by the detector pointing $(\theta_t, \phi_t)$.}

It might seem natural to realise the modulation by $\exp ( - i s \psi_t)$ with an FFT over the pixels of the synthesised maps (Eq.~\ref{eq:f_s}). 
The resulting $f(\psi, \theta, \phi)$ function will have $s_{\mathrm{max}}$ samples over $\psi$ which is typically a low ($s_{\mathrm{max}} \ll \ell_{\mathrm{max}} $) number due to the azimuthal band-limit of the beam.
The TOD can then be directly interpolated from $f$ without manually iterating over $s$. 
In practise, we have found it more efficient in terms of memory and speed as well as more accurate to simply use the $\psi_t$ pointing data and directly apply the factor $\exp(-is \psi_t)$ when the TOD are sampled from the synthesised maps. We thus treat each value of $s$ independently, adding to the TOD with increasing $s$. We use recursion of the form $\exp( i (s + 1) \psi_t)  = \exp( i s \psi_t)  \exp( i \psi_t) $ to avoid unnecessary calls to trigonometric functions.


\subsection{Simulation input}

To evaluate the expression for the beam-convolved data in Eq.~\ref{beam_conv_fast}, we provide the telescope pointing and the spin-weighed spherical harmonic (SWSH) coefficients of the assumed sky and beams. 
We will briefly detail these ingredients.

Following the analytical expression in Eq.~\ref{beam_conv_fast}, the input SWSH coefficient of the beams are the $b^{\widetilde{I}}_{\ell m}$ and ${}_{\mp2} b^{\widetilde{P}}_{\ell m}$ coefficients as defined in Eq.~\ref{blm_I}, \ref{blm_P} and \ref{blm_P_bar}. 
The corresponding beams are assumed to be defined on the $(\theta, \phi)$ coordinate system and should generally be centred on the north pole. 
Because of the redundant description in terms of $\widetilde{P}$ and its complex conjugate, only the $m \geq 0$ modes need to be provided. 
In cases where the co-polar approximation is used (Eq.~\ref{eq:co_pol_diag}) only the $\widetilde{I}$ coefficients are required.

The beam coefficients are associated with one or several detectors. Each detector is represented as a separate instance of a \texttt{Python} class that contains pointers to the beam coefficients as well as properties such as detector pointing offset coordinates, polarization angle and beam band-limits.
Additionally, each detector may be linked to other detector instances that serve as ghosting beams. 
These ghost detectors are treated as fully independent detectors with independent beam coefficients and properties, but are automatically added to the main detector data during data sampling. See the discussion in Sec.~\ref{sec:ghosting}.

Internally, all pointing calculations are performed with \texttt{qpoint} using the computationally efficient unit quaternion representation \citep{Hamilton1866}, rather than the more conventional matrix/vector algebra. 
We separate the pointing information into boresight pointing and per-detector offset pointing coordinates. 
The boresight pointing, representing the pointing direction of the telescope at each time sample, is independent from the  detectors. 
The boresight quaternions (time-ordered data) may either be loaded from disk, calculated in real-time by a user-provided function or one of few preset scanning strategies.
Note that the \texttt{qpoint} library may be used to convert pointing information in Equatorial coordinates (RA, DEC and Position Angle $\psi$) or horizon coordinates (azimuth, elevation and roll) to a suitable time stream of unit quaternions. 

The detector pointing offset is unique to each detector and is assumed to be constant with time. 
The offset physically reflects the different fields of view for detectors placed at different locations away from the telescope's bore axis.
It is realised as an active rotation $g^{(\Delta)}$ away from the boresight pointing direction, specified by an azimuth $a$ and elevation $e$ angle defined relative to the boresight direction (i.e.\ the north pole) $a = e = 0$:
\begin{align}
g^{(\Delta)} = g_{\hat{Z}} (-a) g_{\hat{Y}} (e) g_{\hat{Z}} (0) \quad \quad (\mathrm{ZYZ \: Euler \: convention}) \, .
\end{align}
Here $g_{\hat{Z}}$ and $g_{\hat{Y}}$ represent  rotations around the fixed $Z$ and $Y$ axes respectively.\footnote{Note that we do not include the detector's polarization angle $\gamma$ as a first rotation.
Using $g_{\hat{Z}} (\gamma)$ as the starting rotation in the above would also erroneously rotate the unpolarized beam and its (possibly nonzero) azimuthally asymmetric modes.}  
The polarization angle does not correspond to a physical rotation but is considered as an intrinsic property of the linearly polarized beams and is therefore effectively applied to the $\widetilde{P}$ coefficients. 
The same argument applies to the ideal skyward HWP: its effect is internally handled by modulating the TOD due to the linearly polarized beams by $\exp (\pm 4 i \alpha_t)$ (with HWP angle $\alpha$).

Finally, the harmonic coefficients of the sky are provided in terms of $a^{I}_{\ell m}$ and $E$- and $B$-mode coefficients (see Eq.~\ref{eq:e_b_modes}). Again, only modes with $m \geq 0$ are required.

\subsection{Benchmarks}\label{sec:benchmarks}

We provide some basic benchmark results to illustrate the scaling with beam band-limits $\ell_{\mathrm{max}}$, $s_{\mathrm{max}}$ and the scan duration (see Fig.~\ref{fig:computation_time}). The results consist of two parts: the first shows required CPU time for evaluation of the convolution without any scanning (i.e.\  Eq.~\ref{eq:f_s} for all $ |s| \leq s_{\mathrm{max}}$) as a function of the beam band-limits. The results show the expected total $\mathcal{O}(\ell_{\mathrm{max}}^3 s_{\mathrm{max}})$ scaling. As the azimuthal band-limit will rarely exceed the maximum depicted value $s_{\mathrm{max}} = 8$, the results give a rough indication of wall time in practise. 
The computations are completely dominated by the \texttt{libsharp} inverse SWSH transforms which can be sped up with the use of \texttt{OpenMP} threads and/or \texttt{MPI} tasks (see \cite{reinecke_2013}). For this test we use sequential execution on a single Intel Xeon  E5-2697 v2 core running at 2.7~GHz.

The second result illustrates that sampling the TOD is largely independent of the properties of the beam once the convolved maps (Eq.~\ref{eq:f_s}) are stored in memory. 
This is demonstrated by performing a number of scans with total duration ranging from  $1.5$~min to $50$~days. The sample rate is set at $100$~Hz. The convolved maps and pointing data are preloaded into memory to isolate the test from the inverse SWSH transforms and I/O.
The test is again run sequentially on the same type of core as the previous test. The results are practically identical in case of low resolution ($N_{\mathrm{side}} = 256$) and high resolution ($N_{\mathrm{side}} = 2048$) convolved maps. There is a constant linear scaling with azimuthal band-limit $s_{\mathrm{max}}$. In general, the CPU time for the data sampling part of the procedure scales completely linearly with the number of data samples. 

As expected, the timing results for a completely azimuthally symmetric beam (the black dots) lie approximately a factor $4$ lower than the $s_{\mathrm{max}}=2$ points in the left panel of Fig.~\ref{fig:computation_time}. In this case only $2$ SWSH transforms are used (versus $8$ for the $s_{\mathrm{max}}=2$ case). 

When the two panels in  Fig.~\ref{fig:computation_time} are compared, it can be seen that for small-aperture experiments (i.e.\ $\ell_{\mathrm{max}} \lesssim 2000$) computation time needed for the SWSH transforms is subdominant to that of the data sampling procedure.
As computation time for data sampling remains constant with increasing beam band-limit, the SWSH transforms will dominate computation time for large-aperture telescopes (i.e.\ $\ell_{\mathrm{max}} = \mathcal{O}(10^4)$). We comment on this case in the next section.

\subsection{Future additions}\label{code_future_add}

For high resolution experiments, the multiple inverse SWSH transforms required per detector become  impractical due to their $\mathcal{O}(\ell_{\mathrm{max}}^3)$ scaling. 
When full sky convolution is still desired (in the case of a large observed patch of sky or wide sidelobes), an approach similar to (\cite{Elsner2013}; see also \citep{act_beam_2010, Bicep2_III}) should be used. 
These approaches work by describing the detector beams as linear combinations of a number of basis functions. A small number of basis functions generally suffices due to the relatively small changes in beam properties across a focal plane. 
Each of the basis functions are convolved with the simulated sky map and the resulting maps can be stored in memory shared between computer cores. 
Due to the linearity of the convolution operator, the TOD for each detector can then be sampled from a linear combination of the precomputed maps. 
This approach could even be extended to simulate the effect of detector bandpass differences or beams that are not constant during data acquisition due to e.g.\ temperature drifts or processes dependent on pointing elevation or HWP angle.
We hope to report on the feasibility of this method in future work.

\section{Instrument Setup}
\label{sec:instrument}

\subsection{Overall design considerations}
Using the code library presented in Section \ref{sec:code}, we choose to study a two-lens satellite refractor telescope designed to observe the CMB at two frequencies, 90 and 150 GHz. Designs similar to the one presented here have been considered in design studies for fourth generation CMB satellites \citep{Bock2009, Suzuki2018}. A very rough CAD model is shown in Figure \ref{fig:satellite}. The design includes two silicon lenses embedded in a cold optics sleeve ($\leq 4\,\mrm{K}$) and two concentric radiation shields which prevent direct illumination of the primary lens from the sun. We choose not to incorporate a forebaffle mounted close to the location of the primary lens for fear of polarized reflections and/or increased loading from a blackened load. This puts stringent but quantifiable requirements on internal baffling and scattering in lenses and filters, which would have to be characterised in the lab prior to deployment.

\begin{figure}
\begin{center}
\includegraphics[width=8cm]{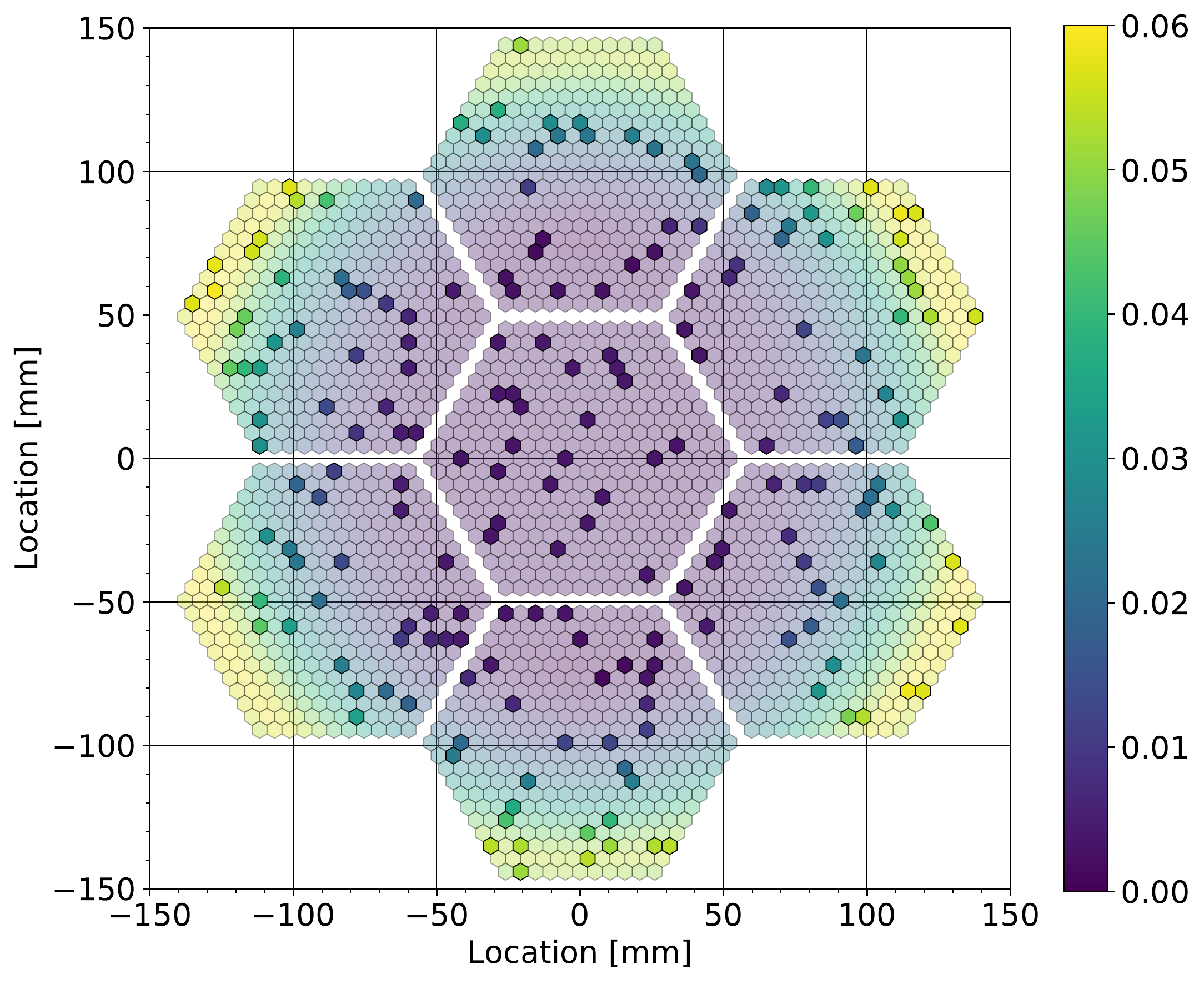}
\caption{Beam ellipticity at 270~GHz as a function of focal plane location. The layout of our proposed focal plane consists of seven detector tiles each with 331 physical pixels of 6 mm diameter, for a total of 2317 pixels. Assuming each pixel is dichroic, this focal plane could support 4634 individual channels. The 200 physical pixels that are included in these simulations are highlighted with zero transparency.  Each physical pixel consists of a pair of orthogonally polarized detectors.}
\label{fig:fpu}
\end{center}
\end{figure}

\begin{figure}
\begin{center}
\includegraphics[width=8.5cm]{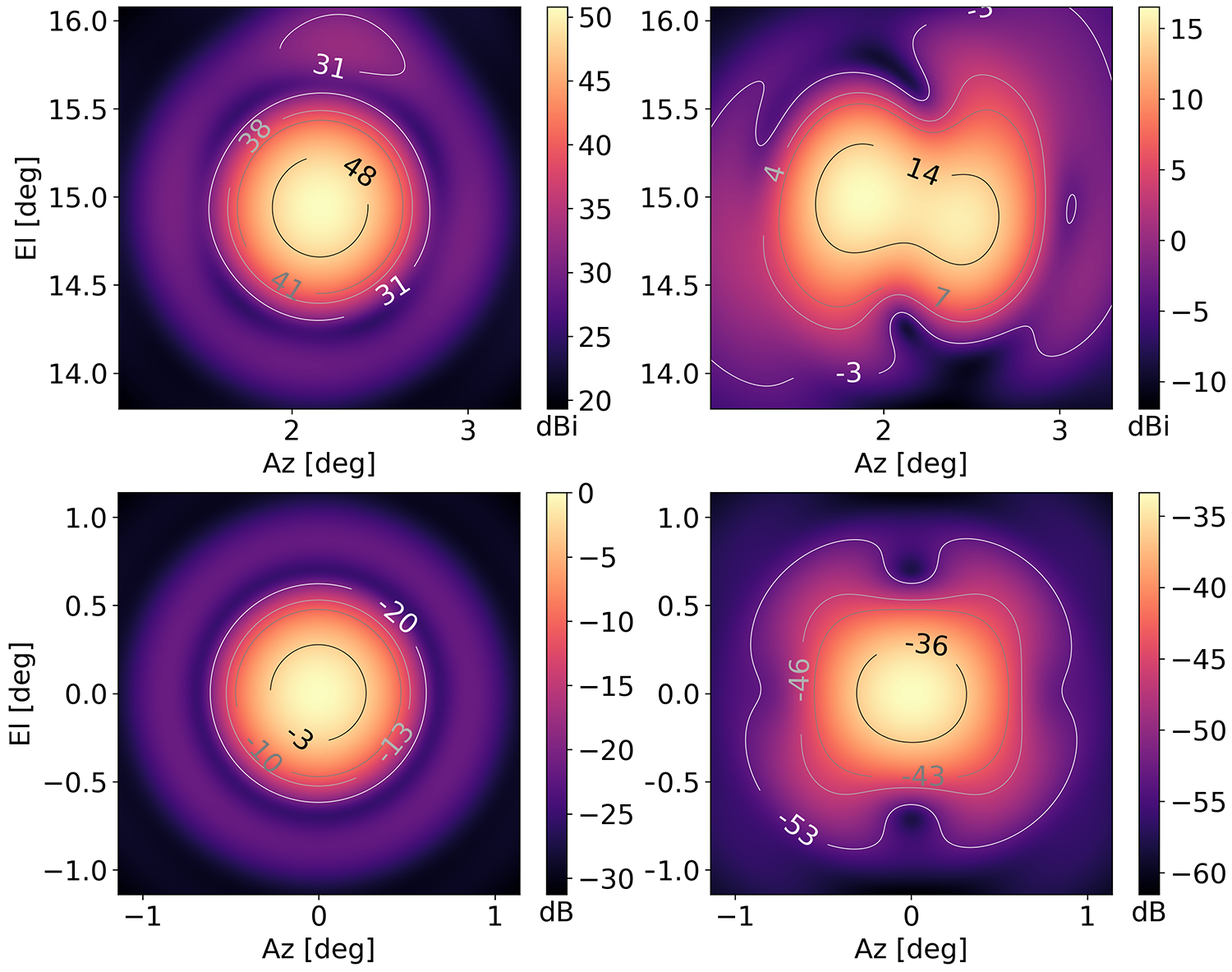}
\caption{Absolute values of beam responses. Top left: Single \hbox{90-GHz} detector co-polar beam map in units of dBi (forward gain over isotropic). The contours indicate -3, -10, -13, and -20~dB relative to maximum, respectively. Top right: The corresponding cross-polar beam map. This particular detector is located approximately 150~mm from the center of the focal plane (edge pixel) and has an ellipticity of $e \approx 0.025$. Bottom left: stacked co-polar beam response at 90~GHz derived by averaging individual beam maps from 100 detectors. Each individual beam is normalized to peak at unity before summing, the final sum is then normalized again. Bottom right: the corresponding cross-polar response which peaks at -33~dB relative to the co-polar response.}
\label{fig:stacked}
\end{center}
\end{figure}

A symmetric on-axis refractor design offers relatively straightforward baffling solutions and a large active focal plane area for a fixed volume design (see Figure \ref{fig:ray_trace}). This design also allows for extensive pre-flight optical characterisation at operational temperatures through the use of a simple test cryostat. Of course, a single optics tube refractor design is hampered by current technological inability to produce anti-reflection coatings that are effective over more than an octave in frequency \cite{Datta2013, Young2017, Defrance2018}. On-axis refractor systems are also more susceptible to internal reflection (ghosting). Although AR-coating challenges of refracting telescopes might mean that reflecting telescopes will ultimately be selected for a 4th-generation CMB satellite mission observing in the primary CMB frequency bands, we choose to further explore this design because of its inherent simplicity and pre-flight characterisation potential.


\subsection{Optical components}
Figure \ref{fig:ray_trace} shows a ray tracing diagram of the proposed two-lens design. The design employs two roughly 380-mm diameter silicon lenses with a maximum zag of about $16\,\mrm{mm}$ on the primary lens. The system has an average effective $f$-number spanning 1.5--1.7 and a telecentricity angle not exceeding $0.1^{\circ}$ over the entire field. The corresponding Strehl ratios at 90, 150, and 270 GHz for this design are shown in Figure \ref{fig:strehl} (see also discussion in Section \ref{sec:po}). Table \ref{tab:opt_prop} describes the key optical design parameters. 

\begin{figure}
\begin{center}
\includegraphics[width=8cm]{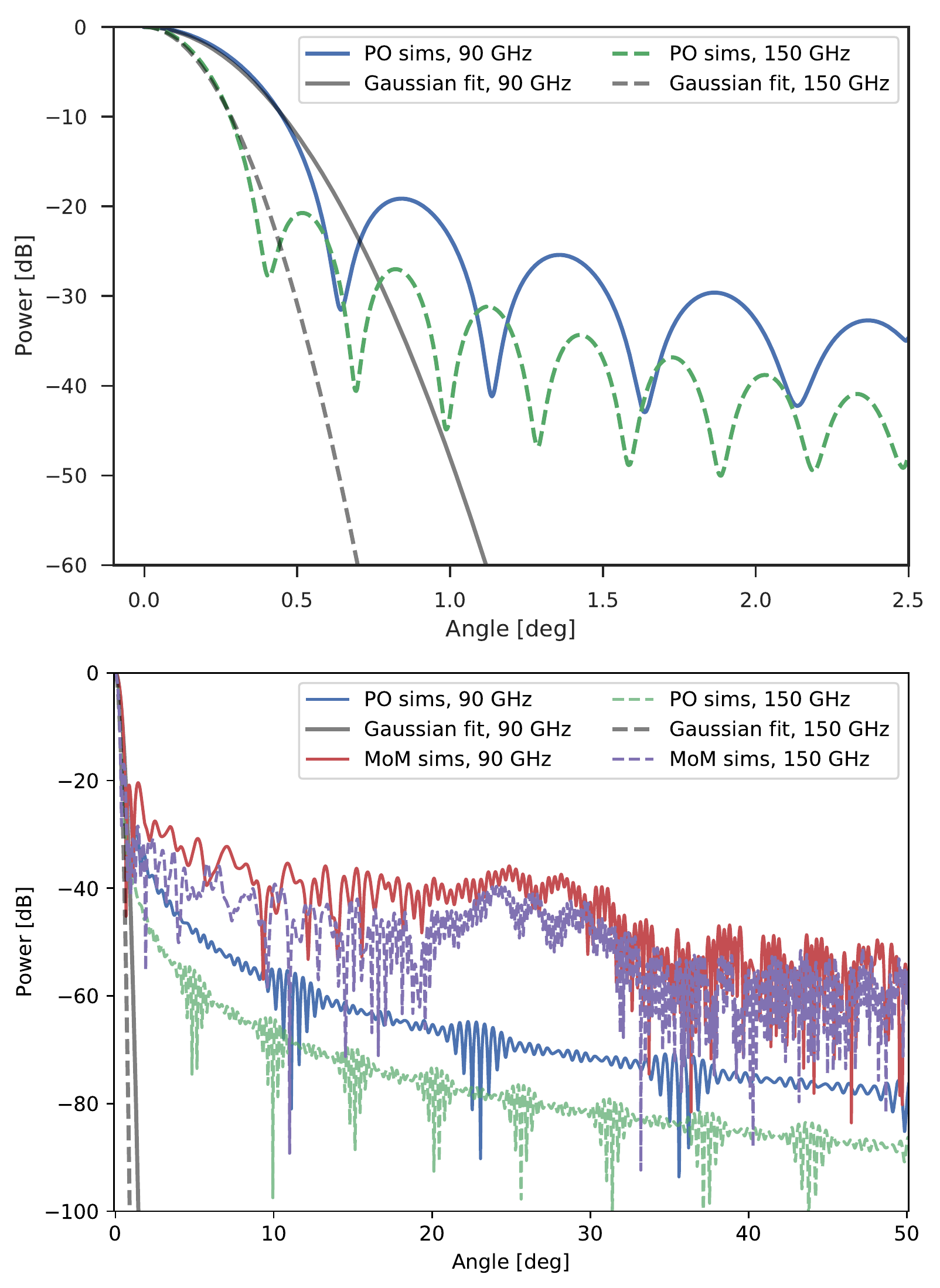}
\caption{Top: Azimuthally averaged beam profiles for the 400 detectors used in these simulations. The best-fit Gaussian beam model to the corresponding stacked (focal plane averaged) beam are shown in grey. Note how the Gaussian model falls off much more quickly with angle. It is clear that significant solid angle is contained in the diffraction sidelobes predicted by GRASP. Bottom: GRASP physical optics (PO) and method of moments (MoM) predictions for the extended sidelobes of the center pixel. Note how the MoM beam profile  has significantly more power at wide angles. This is partially caused by internal reflections in the silicon lenses which are not accounted for by the physical optics calculations. Extended sidelobes can couple to the Galaxy and create a fake polarized signal (see Section \ref{sec:res_sidelobes}). The interference patterns visible in the PO curves (blue and green) are caused by the finite number of frequencies used to simulate the optical response (5 frequencies per band).}
\label{fig:profiles}
\end{center}
\end{figure}

\begin{figure*}
\begin{center}
\includegraphics[width=16cm]{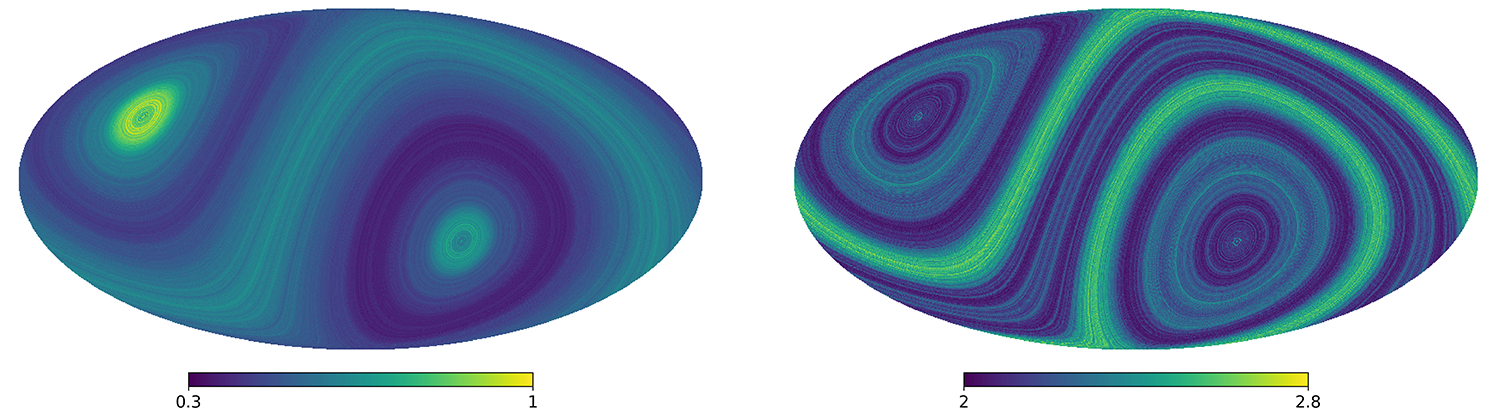}
\caption{Left: Normalized hits map (samples per pixel) in Galactic coordinates for the typical 400-detector scan generated by following the scan strategy presented in Section \ref{sec:scans} for one year at a 96.73~Hz sampling frequency. At this sampling rate, one year of scanning with 400 detectors produces a total of $1.2 \times 10^{12}$ samples. Right: Condition number of $I, Q, U$ covariance matrix for the same scan strategy.}
\label{fig:hitcond}
\end{center}
\end{figure*}


Because of the high index of refraction, the silicon lenses can support a relatively wide diffraction-limited field of view (DLFOV) of approximately 30 degrees. This corresponds to an active focal plane area with a diameter of approximately $290\,\mrm{mm}$ (0.103 deg/mm plate scale) and about 2500 physical pixels, assuming a $6\,\mrm{mm}$ pitch size (see Figure \ref{fig:fpu}). By employing dichroic bolometers with two polarization directions, this telescope could support 5,000 bolometer channels. In comparison, publications discussing the proposed LiteBIRD satellite have suggested that the mission will deploy approximately 2,000 channels \citep{Litebird2016}. A similar number of detectors were proposed for the CORE satellite which employed a two-mirror reflector design \citep{Delabrouille2018}. We note that advances in AR coating technology might allow for the replacement of the centre tile with one populated with pixels spanning the 220- and 270-GHz frequency bands \citep{Coughlin2018, Nadolski2018}. At those frequencies, it would be sensible to deploy smaller pixels to reduce spillover on the cold stop. 

\subsection{Physical optics simulations}
\label{sec:po}
The spatial response of the detectors are simulated using physical optics (PO) and physical theory of diffraction (PTD) simulations as provided by GRASP in results provided by the method of moments (MOM) module \citep{GRASP2018}.\footnote{GRASP is an antenna and optical modeling software capable of providing physical optics and method of moments calculations at mm-wavelenghts. See: \href{https://www.ticra.com/}{https://www.ticra.com/}} The physical optics simulations propagate pixel illumination patterns in succession through the two lenses and out into the far field. The pixel beam illumination pattern is based on a model of a photolithographed bolometer array coupling to corrugated feedhorns, similar to those designed by NIST for ACTPol and Advanced ACTPol \citep{Niemack2010, Koopman2016}. Given the relative simplicity of the optical system, the PO simulations are sufficiently fast that they can be generated for hundreds of detectors in a reasonable amount of time (few days) on a workstation computer. 

In order to capture the focal plane distribution of the beam response, while also providing sufficient coverage to adequately capture aspects of the satellite scan strategy, we have randomly sampled 200 physical pixels spanning the entire focal plane (see Figure \ref{fig:fpu}). In order to inject an additional level of realism to these simulations, we have allowed for some variation in the shape of the pixel beam used to illuminate the secondary lens. The distribution of beam size and ellipticity for the focal plane used in these simulations is shown in Figure \ref{fig:strehl}. We calculate ellipticity, $e$, according to
\begin{equation}
e = \frac{\sigma _x - \sigma_y}{\sigma _x + \sigma _y},
\end{equation}
where $\sigma _x$ and $\sigma _y$ are the Gaussian beamwidths along the two principal axes, with $\sigma _x > \sigma _y$. Figure \ref{fig:stacked} shows the focal plane averaged (stacked) co- and cross-polar beam response at 90~GHz. The stacking procedure washes out any azimuthal asymmetry in individual co-polarized beam maps. Note that the average geometrical cross-polar response is at -33~dB amplitude relative to the co-polar beam. This should be dominated by cross-polar effects originating in the detector architecture itself, for example through cross-talk in detector readout circuits. 


The simulated detector beams are used to create 200 detector pairs. 
This corresponds to a scenario where two perpendicularly linearly polarized radiation coupling devices feed optical power to separate bolometers. 
In this case, the Stokes $\widetilde{I}$ and $\widetilde{V}$ beams are shared between the two bolometers while the $\widetilde{Q}$ and $\widetilde{U}$ beams only differ by a factor $(-1)$ due to the $90^\circ$ polarization angle difference. 
The exact common pointing, shared beams and $90^\circ$ polarization angle difference for all pairs exactly cancels all $I \rightarrow P$ leakage due to differential pointing/beamwidth or azimuthally asymmetric modes of the $\widetilde{I}$ beams (see the discussion in Sec.~\ref{mapmaking}). 
This setup allows us to focus on less explored systematic effects, such as $E \rightarrow B$ leakage due to cross-polar beam components and $m \neq \pm 2$ azimuthally asymmetric modes of the $\widetilde{Q}$/$\widetilde{U}$ beams. 
Of course, this cancellation is only approximate in realistic cases; such modifications could be included trivially in the presented framework. 
For example, in Sec.~\ref{sec:hwp_modulation} we relax this condition by breaking some of the detector pairs to illustrate the $I \rightarrow P $ leakage due to the azimuthally asymmetric modes of the $\widetilde{I}$ beams.

We convert the physical optics results into (spin-weighted) harmonic modes of the corresponding beams following the method explained in Appendix~\ref{app:grasp2jones}. We use a band-limit of $\ell_{\mathrm{max}} = 1000$ and use $m_{\mathrm{max}} = 4$ as azimuthal band-limit for each of the beams. We find the beam components with $m > 4$ too small (on all angular scales) to be significant for the presented analysis.
 

\subsection{Simulation of far sidelobe response}
\label{sec:sidelobe}

The physical optics simulations described in Section \ref{sec:po} naturally incorporate lens and cold stop diffraction effects that cause far sidelobe response. However, those simulations do not factor in the impact of the two radiation shields and/or other passive optical components, such as internal baffling, on the far field response of the telescope. Scattering from impurities and other non-idealities in silicon lenses and filters as well as reflections internal to the optics tube are particularly challenging to model and we omit those effects in the general part of this analysis.

Figure \ref{fig:profiles} shows the 90- and 150-GHz azimuthally averaged beam profiles that are predicted by the physical optics simulations. With the exception of the beams used for analysis presented in Section \ref{sec:res_sidelobes}, the beam profiles predicted by physical optics are apodized at a $4^{\circ}$ angle from beam centre as part of the spherical harmonic decomposition required for \texttt{beamconv} input. Of course, off-axis pickup will continue past this 4-deg cutoff angle. In order to further explore far-sidelobe pickup as a candidate for degree scale $B$-mode systematics, we  look at predictions for sidelobe response from both physical optics and method of moments simulations; these results are also shown in Figure \ref{fig:profiles}. In the reciprocal  sense, the method of moments calculation propagates an electric field emitted at the focal plane through the two silicon lenses and an ideal anti-reflection coating. Surface currents induced in these materials are then combined with the field sourced from the focal plane to calculate a far-field electric field distribution. The off-axis beam response from these method of moments calculations, which is only calculated for a pixel at the centre of the focal plane, is then combined with the physical optics beam maps produced on a per-pixel basis to create a hybrid beam model that is used for the analysis presented in Section \ref{sec:res_sidelobes}. Off-centre pixels will obviously have non-symmetric sidelobes; however, we choose to only conduct a single full-sky MoM calculation in order to save computation time.

As is evident from Figure \ref{fig:profiles}, the sidelobe amplitude predicted from the MoM calculation is significantly higher than that of the PO calculations. This is partially caused by the fact that the method of moments approach ignores passive optical elements such as a cold and absorbing optics tube and allows fields to freely propagate past the primary lens. In comparison, the physical optics calculations effectively ignore any power that does not propagate through the primary lens. We believe that the MoM beam profile response represents a worst case scenario for the proposed optical design. However, we also note that 1\% Lambertian scattering in the silicon primary lens will create a beam sidelobe profile with a comparable amplitude.

\subsection{Input maps}
We use two sets of \texttt{HEALPix} input maps to generate the simulations presented in Section \ref{sec:results}: a CMB-only map generated as a Gaussian random field using the \texttt{synfast} program (part of the \texttt{HEALPix} software library) using a standard $\Lambda \mrm{CDM}$ cosmology, and a map that combines that CMB map with an estimate for dust contributions ($I$ and $P$) in our own galaxy based on a Commander dust foreground template \citep{planck_2015_ix}.

\begin{figure*}
\begin{center}
\includegraphics[width=15cm]{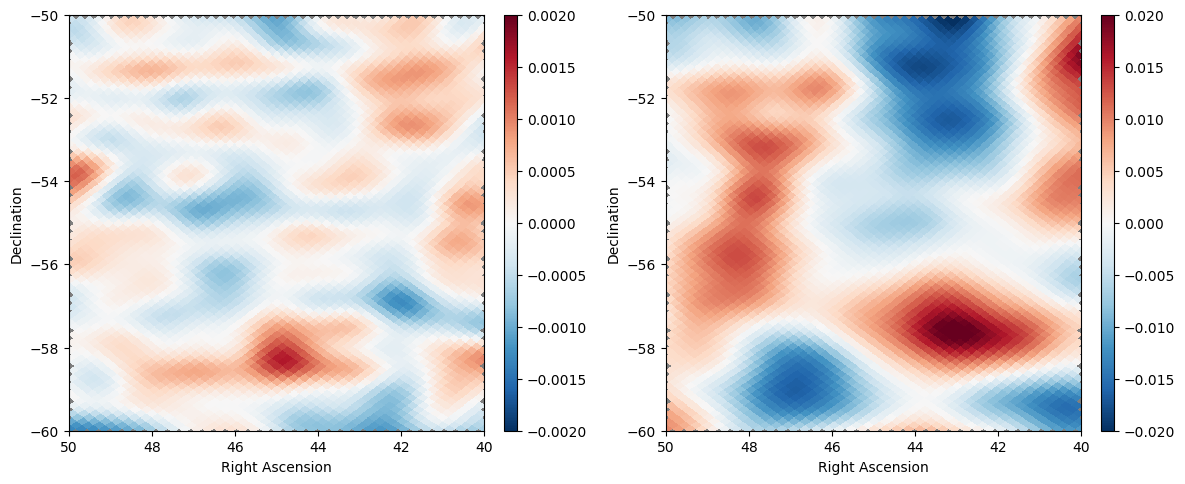}
\caption{Left: A 10-deg square box showing a Stokes-$Q$ difference map between scanning simulations conducted assuming a symmetrical Gaussian beam model and an elliptical Gaussian beam model (units are $\mu$K). Right: Same as left panel, but now differencing a Gaussian beam model with a full physical optics beam model. Note that the color scales on the two panels differ by a factor of 10 and the physical optics model shows substantially more large scale power in addition to the smaller angular scale residuals which have lower amplitude.}
\label{fig:gegpo_diff}
\end{center}
\end{figure*}

The power spectrum input to \texttt{synfast} has neither primordial nor lensing $B$-modes at all scales; this allows us to attribute residual $B$-mode power in rescanned maps to beam non-idealities. The Commander dust template allows us to assess interplay between beams and the galaxy, including the impact of $I \rightarrow P$ leakage through sidelobe coupling to low galactic latitudes. This particular beam-related systematic might constitute a significant challenge for CMB experiments hoping to constrain the epoch of reionization by measuring the associated large-scale $E$-mode signal.

\subsection{Scan strategy, sampling frequency, and duration}
\label{sec:scans}
	
Optical scan strategy for CMB polarimetry has been the subject of multiple publications \citep{Delabrouille2000, Dupac2005, Wallis2017, Natoli2017}. We consider a relatively well-studied satellite scan strategy for L2-observations where the boresight angle, $\beta$, and precession angle, $\alpha$ sum up to approximately 90 degrees (not to be confused with the HWP and boresight rotation angles). The strategy achieves a high degree of cross linking across the entire sky in a full year while maintaining a sun-avoidance angle of greater than $88^{\circ}$ at all times. Using the nomenclature established in \cite{Wallis2017} we set $\alpha=45^{\circ}$ and $\beta=47^{\circ}$, with spin and precession periods of $T_\mrm{spin} = 1\,\mrm{min}$ and $T_\mrm{prec} = 100\,\mrm{min}$, respectively. The scanning strategy and the proposed baffling solution ensure that the primary lens is only illuminated by the sun at glancing angles, if at all. These quantities are summarized in Table~\ref{tab:scanning}.

\begin{table}
\begin{center}
\caption{Satellite scanning parameters and overall beam properties used in these simulations. \label{tab:scanning}}
\begin{tabular}{llll}

\multicolumn{4}{c}{\textbf{Satellite scanning properties}} \\ 
\hline
\multicolumn{3}{l}{Orbit} & L2 \\
\multicolumn{3}{l}{Simulated duration} & 365 days \\
\multicolumn{3}{l}{Sampling frequency, $f_\mathrm{samp}$} & 96.73 Hz \\
\multicolumn{3}{l}{Spin period, $T_\mathrm{spin}$} & 60 s \\
\multicolumn{3}{l}{Precision period, $T_\mathrm{prec}$} & 6000 s \\
\multicolumn{3}{l}{Primary aperture diameter} & 38~cm \\
\hline
\\
\multicolumn{4}{c}{\textbf{Band properties}} \\ 
\hline
Channel & Count & Beam width & Ellipticity  \\ 
{[}GHz{]} & & {[}arcmin{]}&  \\ 
\hline
90 & 400 & 30.4 $\pm$ 0.2 & 0.007 $\pm$ 0.005 \\ 
150 & 400 & 18.9 $\pm$ 0.1 & 0.006 $\pm$ 0.003 \\ 
\hline
\end{tabular}
\end{center}
\end{table}

Figure \ref{fig:hitcond} shows a map of integration time on the sky and condition number of the $\bm{A}^{\dagger} \bm{A}$ matrix per pixel (see Sec.~\ref{mapmaking}) for the scan strategy used in this analysis. This scan strategy uses the 200 bolometer pairs highlighted in Figure \ref{fig:fpu}, with each pixel corresponding to an orthogonally polarized detector pair. The scan strategy is implemented through \texttt{beamconv} interfacing with the publicly available \texttt{qpoint} code. We also use \texttt{qpoint} for the simple binning map-making implemented throughout this paper.
The scanning strategy, detector counts, and sampling frequency result in a relatively Gaussian distribution of condition numbers with an average condition number of $p_\mathrm{cond} = 2.24\pm0.15$ for an $N_\mathrm{side} = 512$ map.\footnote{
The map-maker explicitly solves for $\{\hat{I}, \hat{Q}, \hat{U}\}$ instead of $\{\hat{I}, \hat{P}, \hat{\overline{P}}\}$ but is equivalent to the one described in Sec.~\ref{mapmaking}. It uses $\bm{A} = \left(A_{t,x}\right)^{\mu} \propto \left(1, \cos(2\lambda_t), \sin(2\lambda_t), 0\right) \bm{1}_{X}(t)$ with $\lambda_t = \psi_t + 2\alpha_t + \gamma$ in terms of the position angle $\psi_t$, HWP angle $\alpha_t$ and constant detector polarization angle $\gamma$. The condition number of the (per-pixel) $\bm{A}^{\dagger} \bm{A}$ matrix is defined as the ratio of its largest and smallest singular value and thus has a minimum value of $2$.}  
The full simulation results in $1.2\times10^{12}$ samples per frequency band; each map pixel is therefore visited $3.8\times 10^5$  times in the limit of uniform coverage. See \cite{Wallis2017} for estimates of the expected suppression of leakage per pixel from azimuthally asymmetric beam components for this class of scan strategies.

Optical systematics are tightly coupled to scan strategies, field of view, and sampling frequencies. Therefore, the analysis results presented in Section \ref{sec:results} are only meant to provide qualitative insight. Any real experiment collaboration would have to perform simulations more appropriate for their design.

\section{Results}
\label{sec:results}

Before comparing the results of different beam scanning simulations, we need to standardize the calibration procedures for the different experimental realizations. Here we choose to calibrate the simulations on degree scale temperature anisotropies. Using a temperature power spectrum estimate of the best fit Gaussian beam model for reference, $\hat{C}_{\ell, \mathrm{ref}}^{II}$, we find the best fit scaling parameter, $c$, which minimizes 
\begin{equation}\label{eq:calibration}
\sum _{\ell = \ell_1} ^{\ell_2} R_\ell \equiv \sum _{\ell = \ell_1} ^{\ell_2} \left( c\frac{\hat{C}_{\ell}^{II}}{\hat{C}_{\ell, \mathrm{ref}}^{II}} -1 \right),
\end{equation}
where $\ell_1 = 100$ and $\ell_2 = 300$ and $\hat{C}_{\ell}^{II}$ is a temperature power spectrum calculated using maps obtained from scanning the sky with a more involved beam model. 
Unless the beam models differ significantly from a Gaussian model, this calibration procedure usually results in a number $c \simeq 1$. 
This scaling factor is then applied to all map products and subsequently propagates to all $I$-, $E$-, and $B$-mode power spectra. The power spectrum estimates are calculated using the \texttt{PolSpice} estimator \citep{chon_2004}\footnote{\href{http://www2.iap.fr/users/hivon/software/PolSpice/}{http://www2.iap.fr/users/hivon/software/PolSpice/}}.
We use uniform pixel weighting, except for the Galactic mask used in Sec.~\ref{sec:res_sidelobes}.

A future satellite experiment will likely obtain its absolute calibration on the orbital dipole or cross-calibrate to the temperature anisotropies of past experiments such as \textit{Planck}. The assumed beam model then defines the relative sensitivity of the experiment as a function of angular scale with any error in the beam model propagating to error in the measured power spectra. By choosing degree angular scales ($\ell = 100$--$300$) for our calibration range, we minimize the impact of beam modeling error on the overall amplitude of our derived power spectra.


In the following sections, we study the impact of assuming that the satellite beam model is correctly described by a ensemble average best-fit Gaussian model. We will find that as we increase the complexity of our beam model, the error relative to the Gaussian assumption will grow. Of course, future satellite CMB experiments will calibrate their beam models using point source observations such as planets just as \textit{Planck} and \textit{WMAP} have done in the past \citep{Weiland2011, planckvii_2015, Planck_planets2017}. The beam models constructed in this way will then most likely be supplemented with model predictions from pre-flight measurements as well as ray tracing and physical optics simulations. Any beam model error will then be relative to this more realistic model. In this regard, the comparison conducted in the following sections can be considered that of a worst-case scenario.

\begin{figure}
\begin{center}
\includegraphics[width=8cm]{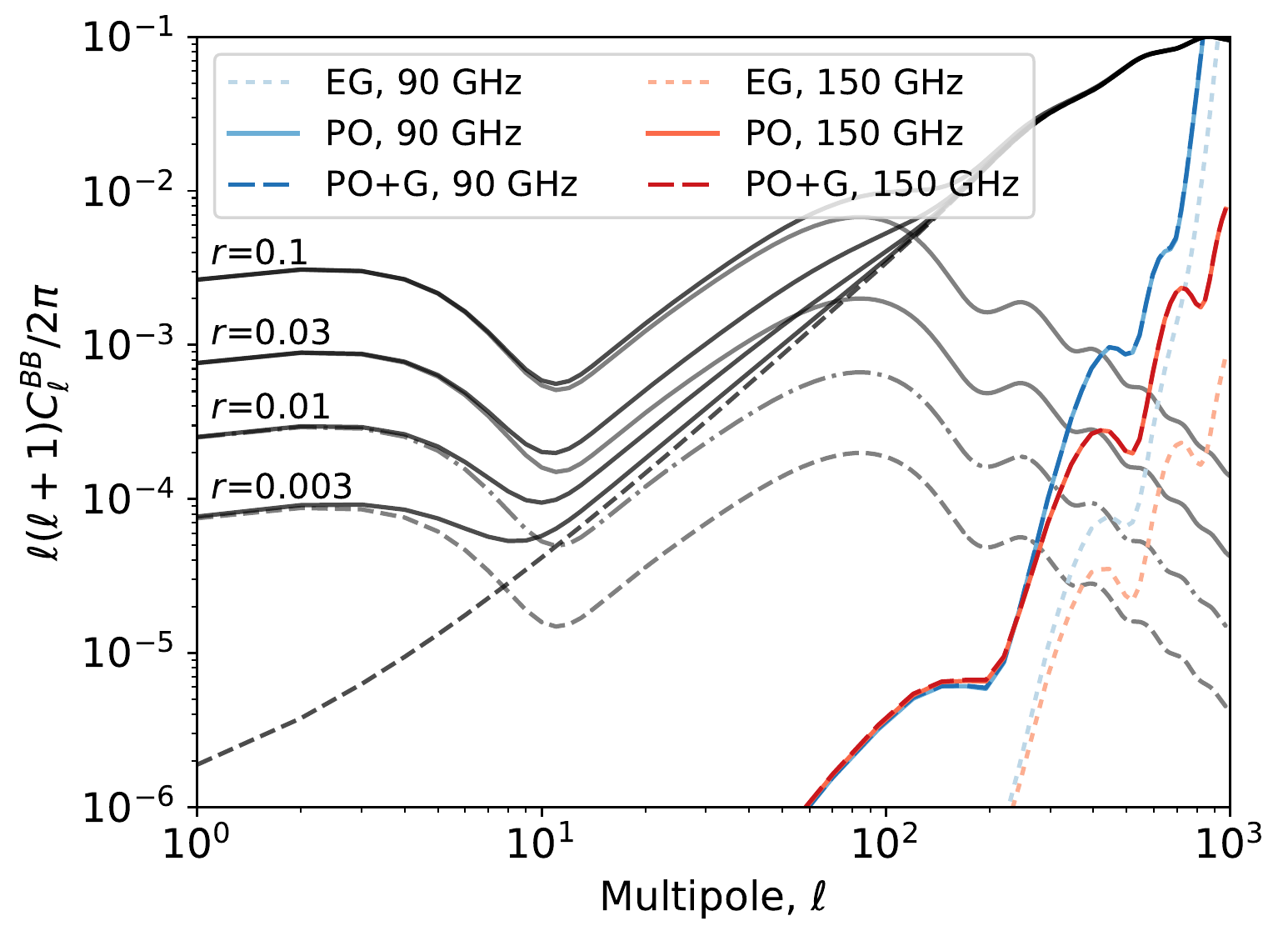}
\caption{Binned 90- and 150-GHz $B$-mode power spectra residuals generated from rescanning the ($B=0$) $\Lambda$CDM map. The two frequencies, 90 and 150~GHz, are shown with blue and red colors, respectively. Three curves, EG, PO, and PO+G correspond to an elliptical Gaussian beam model (dotted curve), a physical optics beam model (solid curve), and a physical optics beam model that includes a ghosting beam response (dashed curve). The larger beam asymmetry at 150 GHz results in a correspondingly larger residual compared to a Gaussian beam model. The grey lines correspond to a primordial $B$-mode power spectrum with a range of tensor-to-scalar ratios while the black solid lines show combination of the primordial and lensing $B$-mode power spectra. Note that the PO and PO+G curves coincide. All power spectra have been deconvolved with an ensemble-average Gaussian beam window function.}
\label{fig:gegpo_spec}
\end{center}
\end{figure}

\begin{figure}
\begin{center}
\includegraphics[width=8cm]{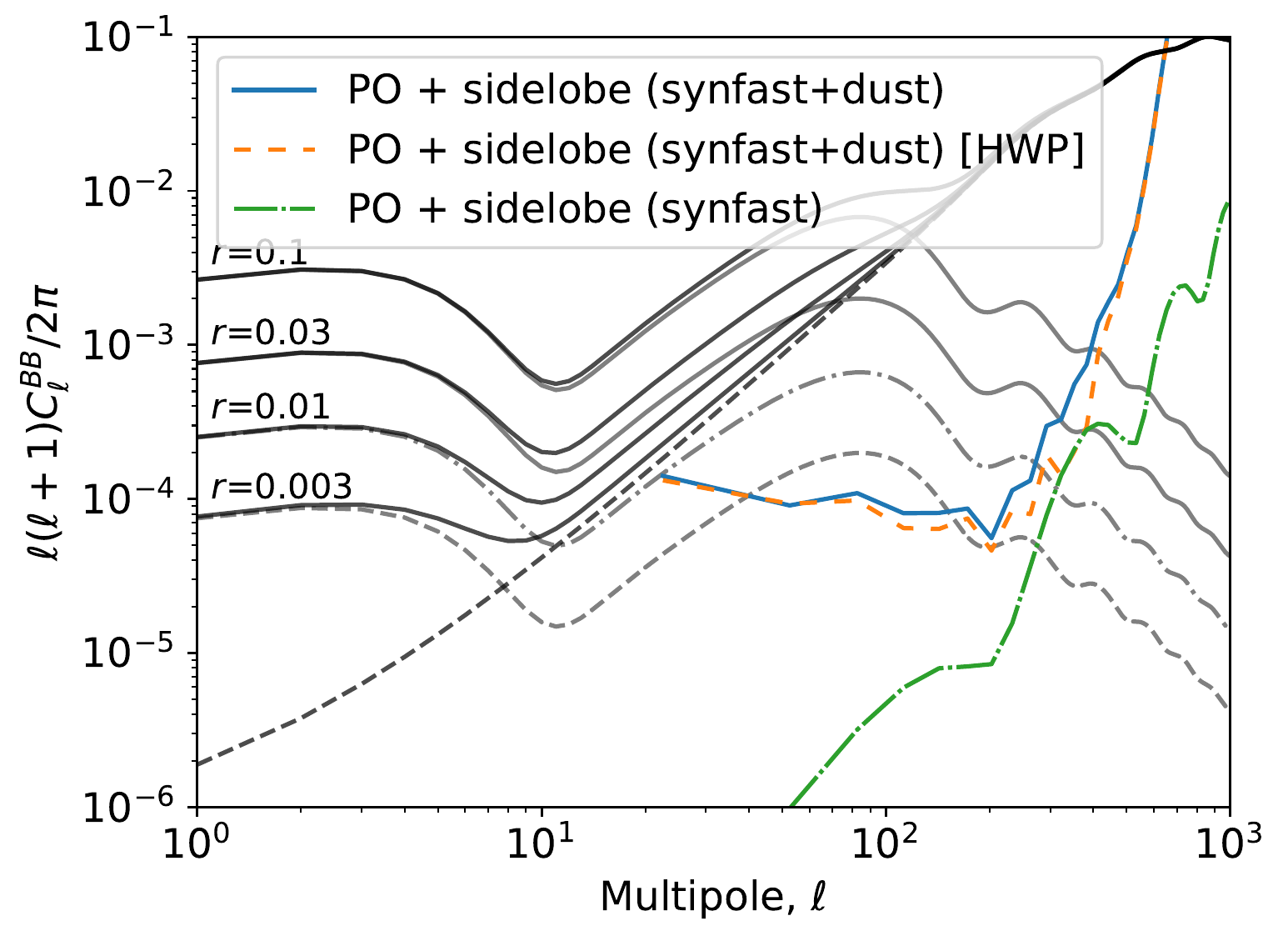}
\caption{Binned 150-GHz $B$-mode power spectra residuals generated from scanning a map formed by combining synfast output with a dust foreground map from Commander (see Section \ref{sec:res_sidelobes}). The solid blue and orange lines correspond to a simulation with and without a continuously rotating HWP, respectively. Low Galactic latitudes are masked using a mask provided by the Planck Collaboration (see Section \ref{sec:res_sidelobes}). The incomplete sky coverage causes oscillations in the PolSpice spectra that are manifest in more ragged power spectra (solid lines). Expected geometric $E \rightarrow B$ leakage due to the mask (not shown) is subdominant over the depicted range in multipole. All power spectra have been deconvolved with an ensemble-average Gaussian beam window function.}
\label{fig:sidelobe_spec}
\end{center}
\end{figure}

\begin{figure}
\begin{center}
\includegraphics[width=8cm]{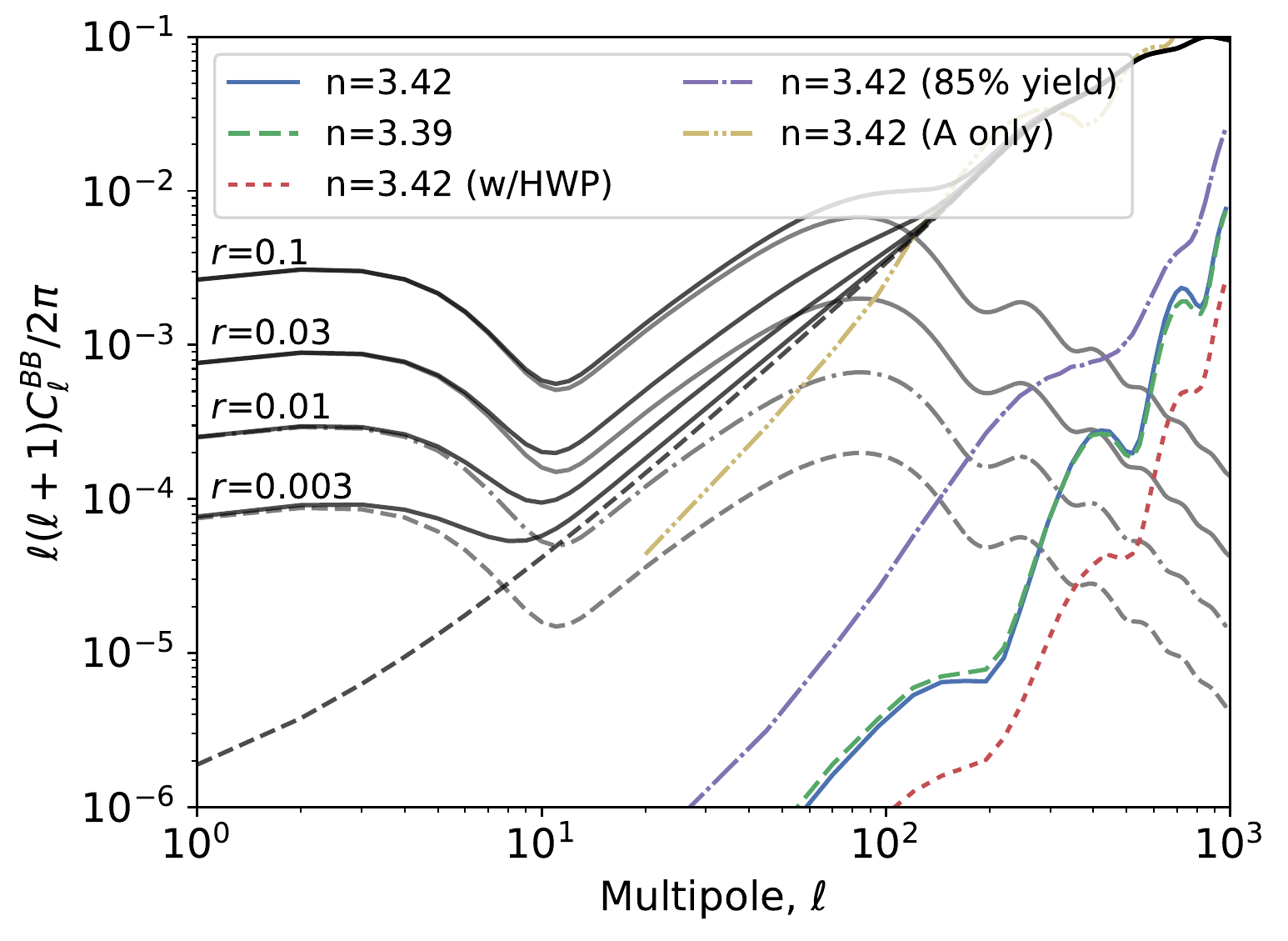}

\caption{Binned 150-GHz $B$-mode power spectra residuals obtained by scanning the ($B=0$) $\Lambda$CDM map with silicon lenses of different index of refraction (see Section \ref{sec:index_tolerancing}). The optics design assumed the $n_\mathrm{si} = 3.42$ (solid blue curve), but we explore a scenario where the effective index is $\tilde{n}_\mathrm{si} = 3.39$ instead (dashed green curve). Note that the solid blue curve here corresponds to the solid red curve in Figure \ref{fig:gegpo_spec}. The dotted red curve corresponds to a scenario where a continuously rotating HWP is placed in front of the primary lens (see Section \ref{sec:hwp_modulation}). The dash-dotted purple line corresponds to a scenario (without a HWP) where a randomly selected 15\% subset of the bolometers are not operating (dead). Finally, the dashed golden line corresponds to a scenario where only a single detector per polarization pair is used for analysis, resulting in maximal $I \rightarrow P$ leakage. All power spectra have been deconvolved with an ensemble-average Gaussian beam window function.}
\label{fig:failed_index}
\end{center}
\end{figure}

\subsection{Gaussian, elliptical Gaussian, and full physical optics beam model}
\label{sec:gegpo}

Simple on-axis optics and high Strehl numbers result in a relatively symmetric beam response across the focal plane (see Figure \ref{fig:strehl}). However, an elliptical Gaussian and a full 2D beam model will both capture the spatial response of this experiment more accurately than a simple Gaussian model. Using \texttt{beamconv} simulations, we can explore the impact of a simple Gaussian assumption at the map level. Before making this comparison, we intercalibrate the different maps using degree-scale temperature cross-correlation (see the discussion around Eq.~\ref{eq:calibration}). This mimics a scenario where the output of these different experiments are calibrated on degree scale power, for example if this experiment were calibrated against the degree-scale power in \textit{Planck} temperature maps instead of deriving absolute calibration on the orbital dipole signal. Absolute calibration on degree scale power likely represents an optimal scenario for experiments focusing on primordial $B$-modes, as this links the calibration to the angular scales of interest.

Figure \ref{fig:gegpo_diff} shows the map-level differences (in $\mu$K) between these different beam model assumptions. It is clear that the physical optics beam model deviates from a Gaussian assumption much more significantly than a simple elliptical Gaussian model (note the different scales on the two color bars). This suggests that an elliptical Gaussian approximation does not fully capture the beam-induced non-idealities in the case of this proposed experiment. It is likely that other experimental setups would produce qualitatively similar outcomes. The corresponding impact on $B$-mode power spectra is shown in Figure \ref{fig:gegpo_spec}. Both the elliptical Gaussian and physical optics beams, when combined with the proposed scan strategy and detector pair matching, create negligible $B$-mode residuals. We also see that the physical optics systematic residual is much greater than the corresponding elliptical Gaussian model residual. This can be attributed to $E \rightarrow B$ leakage from the physical optics beams' increased azimuthal asymmetry and their nonzero cross-polar response.

\subsection{Ghosting beams}

The impact of ghosting beams depends strongly on their amplitude, polarization fraction, and variation across the focal plane. For this publication, we ran a single case where the primary beams are mirrored to diagonally opposite locations on the focal plane with 1\% of the amplitude of the original. In order to add a level of realism, we allow the amplitude of the ghosting beam relative to the main beam to vary by a small amount (relative to 1\%) for every detector. The addition of the ghosting beam does not noticeably increase the beam non-ideality systematic observed in maps and power spectra (see Figure \ref{fig:gegpo_spec}). We expect this simulation to provide a best-case scenario since the shape and amplitude of the ghosting beam for every detector is roughly identical, allowing for significant cancellation through symmetry. In reality, we expect ghosting beams to vary significantly across the focal plane, but the study of that effect is beyond the scope of this paper.

\subsection{HWP modulation}
\label{sec:hwp_modulation}
A rotating half-wave plate reduces susceptibility to low-frequency detector noise which can bias polarization analysis. It also breaks the degeneracy between spurious signal due to azimuthally asymmetric beams and the linear polarized sky signal and therefore reduces potential $I\rightarrow P$ leakage (see Section \ref{sec:hwp_mod}). As a result, some proposed designs for CMB polarimeters include such a device \citep{Litebird2016}. We run simulations with and without an ideal skyward HWP to see how much the $B$-mode power spectrum is impacted. Figure \ref{fig:failed_index} shows the $B$-mode power spectrum differences for the physical optics beam model scanning the sky with and without an ideal HWP spinning continuously at 1~Hz (compare solid blue line with dotted red line). It can be seen that HWP modulation  reduces the $B$-mode residual. Due to the lack of $I \rightarrow P$ leakage, this reduction in power can be attributed to a reduction in $E \rightarrow B$ leakage due to the nearly perfect decorrelation of $Q$ and $U$ in the $\bm{A}^{\dagger} \bm{A}$ matrix from the angular information added by the HWP modulation. The HWP leaves $E \rightarrow B$ leakage due to the cross-polar and asymmetric beam components unchanged, explaining the remaining residual.


\begin{figure*}
\begin{center}
\includegraphics[width=16cm]{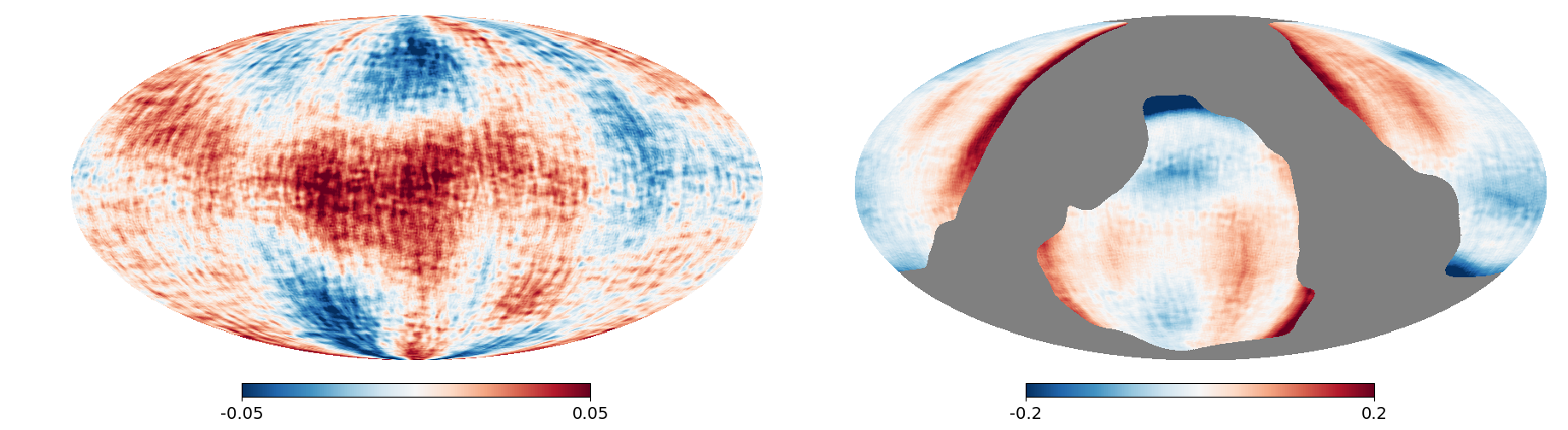}
\caption{Left: Stokes $Q$ difference map (in units of $\mu$K) at 150 GHz as obtained using a wide beam model scanning over a synfast generated map. Right: Same as the left panel, but this time the input map is formed by combining the synfast generated map with a dust template. Note that a difference map with CMB-only map results in even larger residuals. The Planck 60\% mask is included to emphasize residuals away from the Galactic plane. An apodized version of this same mask is used in the calculation of polarization power spectra. It is clear from comparing the two panels that the significant power from the Galactic plane is picked up by the extended beam model. }
\label{fig:sidelobes_diff}
\end{center}
\end{figure*}

All simulations discussed so far have included perfectly orthogonal polarized detector pairs with identical beams for every physical pixel and thus have no contribution from $I \rightarrow P$ leakage due to e.g.\ asymmetric $\widetilde{I}$ beams.
In reality, photolithographed bolometer focal planes suffer from sporadic failures and detector malfunctions. 
These detectors are labelled as dead in the low level analysis and ignored. 
To investigate the effect of non-uniform polarization coverage we randomly suppress 15\% of the 400 detectors that are included in the nominal simulations. This breaks some of the polarization pairs. 
The corresponding $B$-mode polarization residual is represented by the purple dot-dashed line in Figure \ref{fig:failed_index}. 
This loss of pair uniformity clearly increases the residual significantly over the nominal case where all detectors are paired. Taking this to the extreme, we also run simulations with only one detector in each pair, such that all detectors have their polarization sensitivity aligned.
In this case, assuming no spinning HWP, suppression only comes from the angular coverage due to the scan strategy and the experiment is faced with maximal $I \rightarrow P$ leakage (see dashed yellow line). 


\subsection{Adding sidelobes}
\label{sec:res_sidelobes}

Extended sidelobes pick up radiation far from the detector beam centroids. If these sidelobes are sufficiently strong and/or partially polarized this can lead to a significant systematic for CMB polarimeters.

It is particularly interesting to compare the $B$-mode residual for a sidelobe beam model with and without a bright Galactic foreground. Figure \ref{fig:sidelobe_spec} shows the $B$-mode spectra for these two cases. Using a mask that permits 60\% of the sky (see Figure \ref{fig:sidelobes_diff}), the addition of a Galactic dust component significantly increases the level of $B$-mode residual to the point where it is comparable in amplitude to an $r=0.003$ primordial $B$-mode power at degree angular scales. Obviously, this result depends strongly on the mask and adopted sidelobe model; for this analysis we use a standard apodized mask provided by the Planck Collaboration.\footnote{HFI\_Mask\_GalPlane-apo2\_2048\_R2.00.fits} However, we argue that both the mask and the sidelobe model represent realistic scenarios that could apply to future satellite missions. We further note that the addition of a continuously rotating HWP does not ameliorate this systematic since it is not driven by instrument $I \rightarrow P$ leakage.

Figure \ref{fig:sidelobes_diff} shows the 150-GHz Stokes $Q$ difference map (relative to simulation input) for this extended beam model. The beam sidelobe generates large scale residuals regardless of the input; however, the addition of Galactic foregrounds dramatically increases the amplitude of these modes. Because of their complex morphology, Galactic foregrounds coupling to sidelobes will source significant $B$-mode power. We also note that the sidelobes shown in Figure \ref{fig:profiles} are assumed to be completely unpolarized; they are therefore combined with the main beams of both detectors in a polarized pair in an identical fashion. This assumption is incompatible with the fact that sidelobes generated through reflections or diffraction on sharp edges are likely partially polarized. In that sense, assuming identical sidelobes for both detectors in the pair represents an optimistic scenario.

\subsection{Index of refraction tolerancing}
\label{sec:index_tolerancing}

Optical tolerancing typically involves variational analysis that incorporates error in dimensions, locations, and physical properties of different optical components \citep{Page2003, Niemack2016, Parshley2018}. Such errors will lead to an overall deterioration of optical performance, including defocusing and a reduction in optical Strehl ratios. As a simple proof of concept, we chose to study the impact of incorrectly estimating the silicon index of refraction during the design process. We could have just as easily considered changes in the overall shape of one of the lenses, for example due to machining error or unexpected thermal contractions. The principal challenge however, is understanding how design and modeling errors propagate to beam asymmetries and subsequently to primordial $B$-mode residuals.

Figure \ref{fig:strehl} shows the Strehl ratio of the proposed optics design should the effective silicon index of refraction be $\tilde{n}_\mathrm{si} = 3.39$ instead of $n_\mathrm{si} = 3.42$ (see dashed lines). An error of this magnitude is unlikely, but it is instructive to see how the increased beam asymmetry might impact cosmological analysis. Figure \ref{fig:failed_index} compares the $\hat{C}_\ell ^{BB}$ spectra obtained by scanning the sky with physical optics beams obtained using these two different refractive indices at 150~GHz; the impact is less significant at 90~GHz. For these simulations we have implemented the nominal observation strategy which uses 200 bolometer pairs to scan the sky at 96.73~Hz sampling frequency over a duration of one year (see Table \ref{tab:scanning}). From comparing the dashed green curve to the blue curve, we find that the relative importance of such a large modeling error is quite small. This suggests that strong reliance on Strehl ratios as an optical performance metric for 90- and 150-GHz frequency bands might not always be warranted. In particular, there is a relatively small difference between designs with an average Strehl ratio of 0.94 and 0.99 across the entire field of view. Obviously, design choices that cause a 0.05 shift in Strehl ratios at 150 GHz will have an even stronger impact on higher frequencies such as the usual 220- and 270-GHz bands. However, the optical systematics associated with beam asymmetries at those frequencies will be pushed to larger multipoles. In the case of this mock $B$-mode experiment, the primary science goal would not necessarily be affected.

\section{Conclusions}
\label{sec:conclusions}

We have presented and explained the workings of a lightweight, publicly available \texttt{Python} code library capable of generating realistic, full-sky beam-convolved signal timelines that can be subsequently integrated into higher-level simulation pipelines or directly processed into maps and power spectra. 
The code can be used to inform the design of new CMB experiments and characterise existing ones. 

As a proof of concept, we study optical systematics associated with a mock satellite experiment designed to study CMB polarization on degree angular scales. 
As part of this process, we generate realistic estimates for the beam response of the satellite's two-lens refracting telescope that demonstrate relatively large deviations away from the ubiquitous (elliptical) Gaussian beam parameterisation.
In order to focus on several unexplored types of systematic effects, we null the dominant causes for temperature-to-polarization leakage (differential pointing and beam asymmetry). 
We then explore the remaining $E$-mode-to-$B$-mode leakage due to the cross-polar beam components and azimuthal asymmetry of the linearly polarized beam response.
The results indicate that the induced spurious signal is well under control for this setup, even when a deliberate error is introduced in the effective index of refraction of the lenses. 
We note that none of these systematics are negated by the use of a spinning half-wave plate skyward of the primary lens.
Similarly, the addition of ghosting sidelobes due to internal reflections has little effect on the performance of the instrument, but we argue that a more involved study would be appropriate.
The results also highlight the relevance of polarization systematics induced by sidelobe coupling to polarized foregrounds near the Galactic plane.
Finally, we quantify the impact of temperature-to-polarization leakage from reduced detector pair symmetry and demonstrate its problematic nature for a setup without a spinning half-wave plate.

We would like to stress that although these results quantify the amplitude of some optical systematic effects for this setup, general statements are hard to make due to non-trivial dependence on scan strategy and optical design. 
Dedicated simulations are clearly needed for each experimental setup. 
Furthermore, our simulations do not include the effect of telescope components such as filters or baffles, nor do they go beyond the ideal half-wave plate parameterisation. 
Many of these effects can already be included as input to the presented convolution algorithm, but require a more advanced understanding of e.g.\ material properties in optical simulations. 
However, including the effects of non-ideal half-wave plates and their interaction with skyward optical components also requires further development of the convolution algorithm itself. We hope to address some of these questions in future work.

We note that the presented code library is not just capable of simulating systematic effects for $B$-mode power spectrum studies. 
In fact, the method should be especially useful in studies that rely on higher order statistics of the data or studies that directly work with the full dataset (sky maps, or even the time-ordered data). 
We argue that for these sort of studies, forward propagating the optical systematic effects into simulated time-ordered data is the most complete and natural approach to include such effects in the analysis. 
As an example, it would be interesting and timely to investigate how realistic beam effects influence upcoming studies into the CMB polarization field on small-scales over large patches of sky, e.g.\ for lensing estimation.
We suggest a path toward efficiently simulating the full convolution operation for such data and plan to explore such questions in future work.



\section*{Acknowledgements}

We are grateful to Katherine Freese and Martina Gerbino for helpful comments. 
AJD and JEG acknowledge support by Vetenskapsr{\aa}det (Swedish Research Council) through contract No. 638-2013-8993 and the Oskar Klein Centre for Cosmoparticle Physics. JEG acknowledges support from the Swedish National Space Agency (Rymdstyrelsen). Some computations have been performed at the Owl Cluster funded by the University of Oslo and the Research Council of Norway through grant 250672. Some of the results in this paper have been derived using the \texttt{HEALPix} \citep{gorski_2005} package. 




\bibliographystyle{mnras}
\bibliography{paper} 



\appendix
\section{Harmonic representations}
\subsection{Spin weighted spherical harmonic decomposition}\label{app:kernel}

Given generic $\{I, P, V \}$ fields on the sphere, we define the corresponding (spin-weighted) spherical harmonic (SWSH) coefficients as follows (see e.g.~\cite{zaldarriaga_1997}):
\begin{align} 
c^{ {I}}_{\ell m } &= \int_{S^2} \mathrm{d} x\,  {I}(x)  \, \overline{Y}_{\ell m} (x) \, , \label{blm_I}\\ 
{}_2c^{ {P}}_{\ell m } &=  \int_{S^2} \mathrm{d} x\,   {P}(x)  \, {}_2\overline{Y}_{\ell m} (x) \, , \label{blm_P}\\ 
c^{ {V}}_{\ell m }  &=  \int_{S^2} \mathrm{d} x\,  {V}(x) \,  \overline{Y}_{\ell m} (x) \, , \label{blm_V} 
\end{align}
Due to the reality of the Stokes parameters, the $I$ and $V$ coefficients obey $\overline{c^{ {I / V}}_{\ell m }} = c^{ {I / V}}_{\ell -m } (-1)^m$ while the $P$ coefficients obey $\overline{{}_{2}c^{ {P}}_{\ell m }}  = {}_{-2}c^{\overline{ {P}}}_{\ell -m } (-1)^{m}$. The ${}_{-2}c^{\overline{ {P}}}_{\ell m }$ coefficients are the SWSH coefficients of the complex conjugate of $P$:
\begin{align} 
{}_{-2}c^{ {P}}_{\ell m } &=  \int_{S^2} \mathrm{d} x\,   \overline{P}(x)  \, {}_{-2}\overline{Y}_{\ell m} (x) \, . \label{blm_P_bar}
\end{align}
Note that, for brevity, we will write ${}_{-2}c^{ {P}}_{\ell m } $ instead of the more correct ${}_{-2}c^{ {\overline{P}}}_{\ell m } $.

The corresponding forward transforms are given by:
\begin{align}
I(x) &= \sum_{\ell = 0}^{\ell_{\mathrm{max}}} \sum_{m = -\ell}^{\ell} c^{ {I}}_{\ell m } Y_{\ell m} (x) \, , \\
P(x) &= \sum_{\ell = 2}^{\ell_{\mathrm{max}}} \sum_{m = -\ell}^{\ell} {}_{2}c^{ {P}}_{\ell m } \, {}_{2}Y_{\ell m} (x) \, , \\
\overline{P}(x) &= \sum_{\ell = 2}^{\ell_{\mathrm{max}}} \sum_{m = -\ell}^{\ell} {}_{-2}c^{ {P}}_{\ell m } \, {}_{-2}Y_{\ell m} (x) \, , \\
V(x) &= \sum_{\ell = 0}^{\ell_{\mathrm{max}}} \sum_{m = -\ell}^{\ell} c^{ {V}}_{\ell m } Y_{\ell m} (x) \, , 
\end{align}
where $\ell_{\mathrm{max}}$ denotes the band-limit of the field. Note that in the main body of the text we use $\sum_{\ell, m}$ instead of $\sum_{\ell = 0}^{\ell_{\mathrm{max}}} \sum_{m = -\ell}^{\ell}$.

In terms of Euler angles  $(\psi, \theta, \phi)$, we have $\mathrm{d} x = \sin \theta \, \mathrm{d}\theta \mathrm{d} \phi$ and express the spin-weighted spherical harmonics in terms of the Wigner $D^{\ell}$ matrices as:
\begin{align}
D^{\ell}_{-m s } (\phi, \theta, \psi) = (-1)^m \sqrt{ \frac{4 \pi}{2\ell + 1} } {}_s Y_{\ell m} (\theta, \phi) e^{-i s \psi} \, .
\label{D_as_sYlm}
\end{align}
where the coefficients of the $D^{\ell}$ matrices are expressed in terms of the real Wigner-$d^{\ell}$ matrices as:
\begin{align}
D^{\ell}_{m s } (\phi, \theta, \psi) = e^{-im\phi} d^{\ell}_{m s} (\theta) e^{-is\psi} \, .
\end{align}

\subsection{Converting from Ludwig-III to spherical coordinates} \label{app:lud_to_sph}

We may convert the Stokes parameters defined on the Ludwig-III basis to those on the $(\theta, \phi)$ basis as follows:
 \begin{align}
\widetilde{I} &=  \widetilde{I}_{\mathcal{L}} \, ,  \\
\widetilde{P} &=  \widetilde{P}_{\mathcal{L}} e^{- 2i\phi} \, , \label{qu_lud_sph} \\
\widetilde{V} &= -\widetilde{V}_{\mathcal{L}} \, .
\end{align}
We have found that using the above relations to convert beam maps defined on the  Ludwig-III basis to the $(\theta, \phi)$ basis leads to inaccurate harmonic modes of the $\widetilde{P}$ beam. This is ultimately due to the incomplete description of spin-weighted fields on the sphere.\footnote{Recall that a spin-$2$ field, such as $\widetilde{P}$, defined on the tangent space $T_x$ with $x\in S^2$ picks up a factor $e^{-2i\psi}$ under a rotation of the frame through an angle $\psi$ about $x$. Describing this using Euler angles at the pole leads to counterintuitive results, as $\psi$ and $\phi$ become degenerate. A more complete description of the fields we consider would be as functions on the rotation group $SO(3)$, parameterised in terms of unit quaternions (which unlike the Euler angles provide an injective mapping to $SO(3)$ at $\theta=0$).} 
A workaround for this issue is obtained by first calculating the spin-$0$ spherical harmonic coefficients of the $\widetilde{P}_{\mathcal{L}}$ (a well-defined operation at the pole) field and use an analytic expression for the spin-$\pm 2$ spherical harmonic coefficients of the transformation factor $e^{\mp 2i\phi}$ (see \citep{Hivon2017}). By doing so, we may rewrite the above relation for $\widetilde{P}$ in the harmonic domain:
\begin{align}\label{eq:harm_rel_lud}
{}_{\pm2}b^{\widetilde{P}}_{\ell m}  =  \sum_{l'} b^{\widetilde{P}_{\mathcal{L}}}_{\ell' (m\pm2)} K_{\ell \ell' m} \, ,
\end{align}
with spin-$0$ coefficients given by:
\begin{align}
b^{\widetilde{P}_{\mathcal{L}}}_{\ell' m} =  \int_{S^2} \mathrm{d} x\,  \widetilde{P}_{\mathcal{L}}(x)  \, \overline{Y}_{\ell m} (x) \, ,
\end{align}
and the kernel in terms of Wigner-$3j$ symbols:
\begin{align} \label{kernel_expression}
\begin{split}
K_{\ell \ell' m}  = \frac{2}{\sqrt{\pi}} \sum_{\ell''\geq2} & \frac{\sqrt{2 \ell'' + 1}}{\ell'' (\ell''+ 1)} (-1)^{\ell''}
I_{\pm2 0 \mp2}^{\ell \ell' \ell''}  \\ &\times \begin{pmatrix}\ell & \ell' & \ell'' \\ -m & m \pm 2  & \mp 2 \end{pmatrix} \, ,
\end{split}
\end{align}
with:
\begin{align}
I_{\pm2 0 \mp2}^{\ell \ell' \ell''} = \sqrt{\frac{(2 \ell + 1) (2 \ell' + 1) (2 \ell'' + 1)}{4\pi}} \begin{pmatrix}\ell & \ell' & \ell'' \\ \pm 2 & 0  & \mp 2 \end{pmatrix} \, .
\end{align}
The sum over $\ell''$ is formally unbounded. However, the $1/\ell''$ scaling of the factor in front of the $3j$ symbols suppresses any large deviations of $\ell$ from $\ell'$, which in practise means that  $b^{\widetilde{P}_{\mathcal{L}}}_{\ell m}$ and ${}_{\pm2}b^{\widetilde{P}}_{\ell m}$ share band-limits. 

As explained in \cite{Hivon2017}, for a sufficiently localised beam, the kernel $K_{\ell \ell' m}$ may be approximated as diagonal per azimuthal mode $m$, i.e.\ $K_{\ell \ell' m} \approx \delta_{\ell \ell'}  \, \forall m$. We have used this approximation for the presented analysis.

\subsection{Azimuthally symmetric beams}\label{app:az_symm}
Naively demanding that the harmonic modes for the $\widetilde{P}$ beam placed on the pole should also be zero for $m\neq0$ leads to the unphysical result that the beam must vanish at the pole as $\widetilde{P}|_{\theta=0} \propto {}_{2}Y_{\ell m}|_{\theta=0} \propto \delta_{m2}$. 
 As noted before, a way toward a correct expression would be to explicitly represent $\widetilde{P}$ as a scalar field on $SO(3)$ using e.g.\ the unit quaternions, but a simpler approach is to make use of the method presented in Appendix~\ref{app:lud_to_sph}. 
 
We start by placing the (localised) azimuthally symmetric $\widetilde{P}$ beam on the pole, i.e.\ centred around the $\hat{z}$ axis. The $\widetilde{P}$ beam, with Stokes parameters defined with respect to the $(\theta, \phi)$ coordinate system, is related to the same beam defined on the  Ludwig-III basis ($\widetilde{P}_{\mathcal{L}}$) through the relation in Eq.~\ref{qu_lud_sph}. We use this relation to rewrite the beam as $\widetilde{P}_{\mathcal{L}} e^{- 2i\phi}$. We may consider $\widetilde{P}_{\mathcal{L}}$ as a spin-$0$ field, as long as $e^{- 2i\phi}$ is considered a spin-$2$ field (to ensure that the product keeps the correct transformation properties). We have seen that the only nonzero spin-$0$ harmonic modes of $\widetilde{P}_{\mathcal{L}}$ will be those with $m=0$ (see Eq.~\ref{eq:unpol_sym}). Inserting these coefficients, i.e.\  $b^{\widetilde{P}_{\mathcal{L}}}_{\ell' m} \propto \delta_{m0}$, into Eq.~\ref{eq:harm_rel_lud} demonstrates that ${}_{\pm2}b^{\widetilde{P}}_{\ell m}  \propto \delta_{m\mp2}$.

\section{Estimating beams with optical simulations} \label{app:grasp2jones}

As mentioned in the main text (Sec.~\ref{sec:co_cross_beams}), optical simulations may be used to estimate the instrumental response when it is completely described by its co- and cross-polar response.\footnote{In this regime one is unable to probe the depolarizing properties of the instrument as this would additionally require the response to an unpolarized source.} The instrument is then described as a non-depolarizing transformation (excluding the perfectly depolarizing incoherent detector for now); we may either use the Jones or Mueller-Jones transformations to describe such a system. For conciseness we pick the Jones formalism. 

Following \citep{rosset_2010}, we describe the Jones matrix of the instrument as an imperfect linear polarizer coupled to a generic Jones matrix describing the telescope:
\begin{align} \label{eq:imperfect_pol_jones}
J_{\mathcal{L}}(x, \omega) = \begin{pmatrix} J_{11} & J_{12} \\ \sqrt{\eta} J_{21} &  \sqrt{\eta} J_{22} \end{pmatrix} (x, \omega) \, ,
\end{align}
where $\eta \rightarrow 0 $ in case the linear polarizer becomes perfect. $J$ is defined on the Ludwig-III basis, as indicated by the subscript ${}_{\mathcal{L}}$.  The co- and cross-polar responses are due to the $J_{11}$,  $ \sqrt{\eta} J_{21}$ and $J_{12}$,  $ \sqrt{\eta} J_{22}$ elements respectively. Note that the parameter $\eta$ is sometimes referred to as the cross-polar leakage. This label might cause confusion as both the co- and cross-polar response depend on $\eta$. 

In the $\eta = 0$ case, incident co-polar radiation ($\epsilon = \begin{psmallmatrix} 1 \\ 0 \end{psmallmatrix}$) probes the $J_{11}$ component while incident cross-polar radiation probes the $J_{12}$ component. Optical simulations like those described in Sec.~\ref{sec:po}, generally work in the reciprocal sense: they simulate the propagation of electric fields emitted from the location of the detector through the optical system out to the sky. As long as the system is reciprocal, the simulation results may be used to estimate the instrument response to radiation with reversed direction of propagation. In terms of Jones matrices, the reciprocal Jones matrix for such a system is given by $\bm{J}^{\mathrm{recpr.}} = \mathrm{diag}(1, -1) \bm{J}^T \mathrm{diag}(1, -1)$ \citep{sekera_66}. This implies that the nonzero elements of the instrumental Jones matrix may be estimated from the far-field response of electric fields with known amplitude and polarization state that propagated through the system. For example, by emitting a purely co-polar field from the detector, the far-field vector components are proportional to the $\begin{psmallmatrix} J_{11} \\ - J_{12} \end{psmallmatrix}$ elements from Eq.~\ref{eq:imperfect_pol_jones}. Probing the $J_{21}$ and $J_{22}$ components in case where $\eta \neq 0$ amounts to repeating the simulation with a purely cross-polar initial field. For the examples presented in this work we have omitted this last step: assuming perfect linear polarizing elements at the end of the optical chain.

With the instrumental Jones matrix estimated, the last steps are to incorporate the depolarizing incoherent detector and to calculate the three fields that describe the instrumental beam: $\{\widetilde{I},\, \widetilde{P},\, \widetilde{V}\}$. One way to achieve both is to convert the Jones matrix to the associated Mueller-Jones matrix and setting all the elements but those in the top row to zero.\footnote{The associated Mueller-Jones matrix is related to the Jones matrix by: $M^{\mu}_{\phantom{a} \nu}  
=  \frac{1}{2} \mathrm{Tr} \{\bm{\sigma}^{\mu} \bm{J} \bm{\sigma}_{\nu} \bm{J}^{\dagger} \}$ with $\bm{\sigma}_{\mu} = \{ \bm{1}, \bm{\sigma}_{3}, \bm{\sigma}_{1}, \bm{\sigma}_{2} \}$ in terms of the Pauli matrices.} After conversion to the spherical coordinate system, the remaining elements are then equal to the $\{\widetilde{I},\, \widetilde{Q}, \, \widetilde{U},\, \widetilde{V}\}$ beams in the instrument frame and have harmonic coefficients that can be used as input to Eq.~\ref{beam_conv_fast}.
	


\bsp	
\label{lastpage}
\end{document}